\documentclass[10pt,twocolumn,letterpaper]{article}

\usepackage{iccv}              % To produce the CAMERA-READY version
\definecolor{iccvblue}{rgb}{0.21,0.49,0.74}
\usepackage[pagebackref,breaklinks,colorlinks,allcolors=iccvblue]{hyperref}
\usepackage{multirow}
\def\paperID{4214} % *** Enter the Paper ID here
\def\confName{ICCV}
\def\confYear{2025}

\title{Feature Coding in the Era of Large Models: \\Dataset, Test Conditions, and Benchmark}

\author{
Changsheng Gao$^1$, Yifan Ma$^2$, Qiaoxi Chen$^2$, Yenan Xu$^2$, Dong Liu$^2$, Weisi Lin$^{1}$\textsuperscript{\textdagger}\\
$^1$Media Interactive Computing Lab, Nanyang Technological University, Singapore 639798\\
$^2$MOE Key Laboratory of Brain-Inspired Intelligent Perception and Cognition,\\
University of Science and Technology of China, Hefei 230093, China\\
\{changsheng.gao, wslin\}@ntu.edu.sg; \{mayf,xxii,yenanxu\}@mail.ustc.edu.cn; dongeliu@ustc.edu.cn\\
}

\begin{document}
\maketitle
\renewcommand{\thefootnote}{}
\footnotetext{This work was supported by the Ministry of Education of Singapore under Grant T2EP20123-0006. We acknowledge the support of GPU cluster built by MCC Lab of Information Science and Technology Institution, USTC.}
\begin{abstract}
Large models have achieved remarkable performance across various tasks, yet they incur significant computational costs and privacy concerns during both training and inference. Distributed deployment has emerged as a potential solution, but it necessitates the exchange of intermediate information between model segments, with feature representations serving as crucial information carriers. To optimize information exchange, feature coding is required to reduce transmission and storage overhead. Despite its importance, feature coding for large models remains an under-explored area. In this paper, we draw attention to large model feature coding and make three fundamental contributions. First, we introduce a comprehensive dataset encompassing diverse features generated by three representative types of large models. Second, we establish unified test conditions, enabling standardized evaluation pipelines and fair comparisons across future feature coding studies. Third, we introduce two baseline methods derived from widely used image coding techniques and benchmark their performance on the proposed dataset. These contributions aim to provide a foundation for future research and inspire broader engagement in this field. To support a long-term study, all source code and the dataset are made available at \href{https://github.com/chansongoal/LaMoFC}{https://github.com/chansongoal/LaMoFC}.

\end{abstract}
\begin{figure*}[tp]
	\centering
	\includegraphics[width=0.98\linewidth]{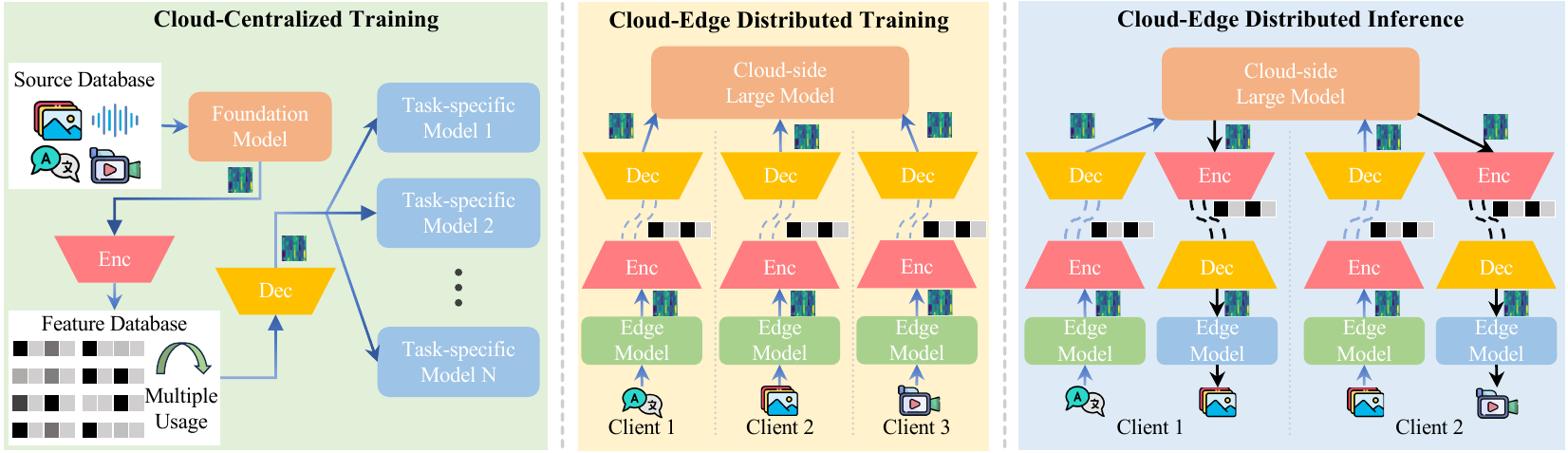}
    \caption{Three application scenarios for \textbf{feature coding} in large model deployments. \textbf{Cloud-centralized training:} Features generated by foundation models are encoded and stored in a feature database. This database supports various downstream tasks, reducing computational demands by avoiding repeated inference of large foundation models. \textbf{Cloud-edge distributed training:} To protect user data privacy, raw data is first processed by an edge model to generate features, which are then encoded and transmitted to the cloud. The cloud decodes these features to complete the remaining forward propagation. \textbf{Cloud-edge distributed inference:} A large model is partitioned across multiple parts, with portions offloaded to edge devices to distribute the computational load. The transmitted features ensure privacy protection and enable customized downstream tasks. Please refer to Sec. \ref{sec_scenario} for detailed illustrations.}
	\label{fig_scenario}
\end{figure*}
\section{Introduction}
\label{sec_introduction}
Large models have revolutionized artificial intelligence (AI) with their impressive performance. However, they require substantial training data and extensive computational resources, posing significant challenges for practical deployments \cite{kaddour2023challenges}. To address these challenges, distributed deployment has been proposed as a promising solution \cite{tian2022fedbert,vepakomma2018split,ye2024openfedllm,zheng2024safely,lepikhin2020gshard,friha2024llm}. By distributing the computational workload across multiple platforms, this approach reduces the resource strain on individual systems and enables the integration of private client data into training while maintaining data security \cite{ye2024openfedllm,chen2024federated}. As model scale continues to grow, distributed deployment is expected to become a mainstream strategy for large model deployments.

In distributed deployments, effective information exchange between model segments is essential. During forward propagation, features generated by early segments are passed to later segments to complete the inference. Given the vast amount of data in large model applications, it is necessary to encode features to a smaller size to reduce storage and transmission burden. Analogous to image and video coding, feature coding has been proposed to address this problem \cite{chen2020toward,gao2024imofc,gao2024dmofc,wang2022towards}.

Existing feature coding research faces three key limitations: narrow model types, limited task diversity, and constrained data modalities. Most existing feature coding efforts focus on small-scale convolution-based networks, such as ResNet, ignoring large models. 
% Although existing research emphasizes discriminative models, large models such as DINOv2 \cite{oquab2023dinov2} and SAM \cite{kirillov2023segment} remain underrepresented. Generative models also have received little attention in this domain. 
Furthermore, existing research primarily targets visual features and neglects the investigation of textual features from large language models (LLMs). This gap indicates a need for broader exploration across both models and data modalities to advance feature coding.

In this work, we shift the feature coding research from small-scale to large-scale models. To address the existing limitations, we construct a comprehensive test dataset with features from five tasks across three large models, covering both visual and textual data for increased modality diversity. Furthermore, we establish unified test conditions to enable fair comparisons, including formulated bitrate computation and task-specific evaluation pipelines.

To lay the groundwork for large model feature coding, we propose two baseline methods and evaluate their applicability to the dataset. The baselines leverage two image coding methods: the handcrafted VVC Intra coding \cite{bross2021vvc} and the learning-based Hyperprior method \cite{balle2018variational}. We introduce pre-processing and post-processing modules to adapt the baselines to feature characteristics. 
The resulting benchmark and analyses provide valuable insights into large model feature coding and identify promising directions for further exploration.
In summary, our main contributions are as follows:
\begin{itemize}    
    % \item We highlight the importance of feature coding in large model deployment, identifying relevant application scenarios and extending feature coding research from small- to large-scale models.
    \item We highlight the importance of feature coding in large model deployments, identify relevant application scenarios, and introduce a new research area: large model feature coding. 
    \item We construct a comprehensive feature dataset encompassing three model types, five tasks, and two kinds of source data modalities, along with unified test conditions to support long-term large model feature coding research.
    % \item We propose and benchmark two feature coding methods on the dataset, offering baseline evaluations and insights for future research in this field.
    \item We propose two baselines and evaluate their usability to the large model feature coding. We build a benchmark and suggest promising directions for further research.
\end{itemize}

\begin{table*}[tbp]
	% \footnotesize
	\centering
    \resizebox{\textwidth}{!}{
	\begin{tabular}{@{}c|c|c|c|c|c@{}}
		\toprule
		\textbf{Model}                   & \textbf{Parameters}   & \textbf{Task}         & \textbf{Split Point}              & \textbf{Feature Shape} & \textbf{Source Data}         \\ \midrule
		\multirow{3}{*}{\textbf{DINOv2}} & \multirow{3}{*}{1.1B} & Image Classification (Cls) & $40^{th}$ ViT Block                    & $257\times1536$               & ImageNet, 500 samples        \\
		&                       & Semantic Segmentation (Seg) & $40^{th}$ ViT Block                    & $2\times1370\times1536$            & VOC2012, 100 samples      \\
		&                       & Depth Estimation (Dpt)     & $10^{th}$, $20^{th}$, $30^{th}$, $40^{th}$ ViT Blocks & $2\times4\times1611\times1536$          & NYU-Depth-v2, 80 samples            \\ \midrule
		\textbf{Llama3}                 & 8B                    & Common Sense Reasoning (CSR)      & $32^{th}$ Decoder Layer                & $N\times4096$                 & Arc-Challenge, 500 samples   \\ \midrule
		\textbf{SD3}      & 8B           & Text-to-Image Synthesis (TTI)       & Input Layer of VAE Decoder                  & $16\times128\times128$             & COCO2017, 500 samples \\ \bottomrule
	\end{tabular}}
    \caption{Summary of the model and task selection, split point decision, and source data collection.}
    \label{table_summary}
\end{table*}
\section{Background and motivation}
\label{sec_background} 
\subsection{Feature coding}
Originally introduced within the Coding for Machines framework \cite{yang2021video,duan2020video}, feature coding is one of two primary branches alongside visual coding. Unlike visual coding \cite{tian2023nonsemantics,zhang2024all,gao2023towards,lu2024preprocessing,tian2024smc,mao2024perceptual,sheng2024vnvc,li2024object,li2024ustc}, which encodes and reconstructs the original visual data, feature coding involves encoding features extracted from source data into bitstreams. While visual coding has already been explored in the context of large models \cite{tian2024freevsc,kao2024bridging}, feature coding remains unexplored in this domain.

Despite recent progress, feature coding research still faces three key limitations. First, there is a \textbf{restricted variety of model types} in use. 
The existing feature coding focuses on small-scale CNN features such as \cite{li2023attention,suzuki2022deep,kim2023end,liu2023learnt,cai2022high,ma2024feature,gao2024rethinking}. However, as Transformer architectures now dominate large model research \cite{touvron2023llama,dubey2024llama,esser2024sd3,oquab2023dinov2}, it is crucial to investigate feature coding for Transformer-based models.
Second, the field is constrained by the \textbf{limited scope of task types}. Existing studies primarily address discriminative tasks such as image classification and segmentation \cite{choi2021latent,feng2022image,zhang2021MSFC,yan2021SSSIC,chen2024end,misra2022video}, while generative tasks remain largely unexplored. Given the significance of generative models, extending feature coding to include these tasks, such as image synthesis, is essential.
The third limitation is the \textbf{constrained range of source modality}. 
Current feature coding research predominantly focuses on features extracted from visual data, overlooking the increasingly important features generated from textual data, particularly in light of advancements in LLMs. Expanding to multiple modalities is crucial as the diversity of AI applications grows.

Furthermore, two challenges hinder progress in large model feature coding: \textbf{the lack of a public test dataset and unified test conditions}. Without a standardized dataset and evaluation conditions, fair comparisons across methods can not be conducted. A public dataset ensures consistent input data across studies, while unified test conditions establish a standard pipeline to evaluate performance.

These limitations motivate us to construct a diverse test dataset and formulate unified test conditions to support large model feature coding research.

\subsection{Large model deployment}
The impressive performance of large models is driven by scaling laws \cite{radford2021learning,sun2023eva}, which emphasize the importance of vast model sizes and extensive training data. For example, GPT-3 was built with 175 billion parameters and trained on 499 trillion tokens, both contributing to substantial computational demands. To alleviate these resource requirements, distributed training and inference methods have been proposed in \cite{ye2024openfedllm,zheng2024safely,friha2024llm}.

User privacy protection presents another challenge in client services \cite{chen2024federated,lyu2024privacy,yan2024protecting}. Split learning has been proposed as an effective solution to protect data privacy by partitioning models between clients and the cloud \cite{ye2024openfedllm}.

In distributed deployments, efficient information exchange between model segments is crucial. However, academic research has largely overlooked the storage and transmission costs associated with feature exchange. To address this gap, we propose to encode features into compact bitstreams to improve the efficiency of large model deployments in distributed environments.

\section{Feature coding in large models}
\label{sec_scenario}
In Fig. \ref{fig_scenario}, we illustrate three application scenarios of feature coding in large model deployments, highlighting their roles in optimizing the deployment efficiency of large models.
\subsection{Large model training}
Large models achieve their high performance through extensive training on large-scale datasets. Cloud-centralized training is commonly employed to leverage the vast computational resources and data in the cloud. However, in sensitive domains like healthcare, high-quality public data may be scarce, necessitating access to private data through cloud-edge collaborative training. 
\subsubsection{Cloud-centralized training}
In cloud-centralized training, source databases are maintained in the cloud, where intensive computational demands become the main challenge. Complex tasks are increasingly addressed through the collaborative use of multiple models. For instance, in vision-language tasks, visual data is first processed by a vision model, and its features are then passed to an LLM for advanced reasoning. Similarly, foundation models often serve as core backbones for various tasks. For example, features extracted from DINOv2 are repurposed for downstream tasks like segmentation \cite{docherty2024upsampling}. In such collaborative workflows, features generated by one model become inputs for others. However, during training, each data sample may undergo repeated inference, resulting in substantial and unnecessary computational costs.

To address this, an effective solution is to generate features once, store them, and reuse them across tasks. The workflow is outlined on the left of Fig. \ref{fig_scenario}. First, a foundation model extracts features from the source database, which are then encoded and stored in a feature database. For subsequent task-specific training, only the precomputed feature database is needed, enabling efficient access to features without redundant feature extraction. This approach significantly reduces computational load and accelerates downstream task training.
However, raw feature data consumes considerable storage space, particularly with large datasets. Therefore, encoding raw features into compact bitstreams becomes essential to reduce storage demands and support efficient large model training.

\begin{figure*}[tbp]
	\centering
	\begin{minipage}{0.45\columnwidth}
		\centerline{\tiny{$SP_{DM1}$}}
		\vspace{0.2em}
		\includegraphics[width=\linewidth]{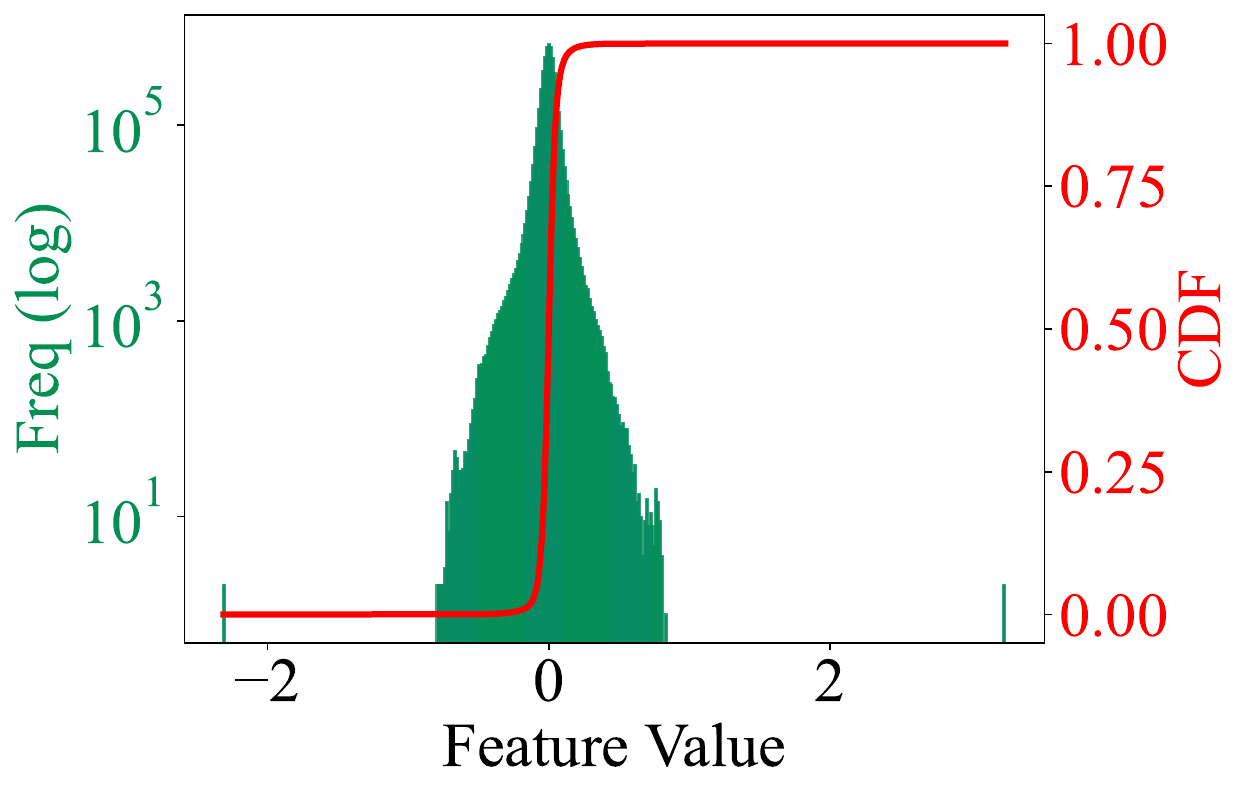}
	\end{minipage}
	\begin{minipage}{0.45\columnwidth}
		\centerline{\tiny{$SP_{DM2}$}}
		\vspace{0.2em}
		\includegraphics[width=\linewidth]{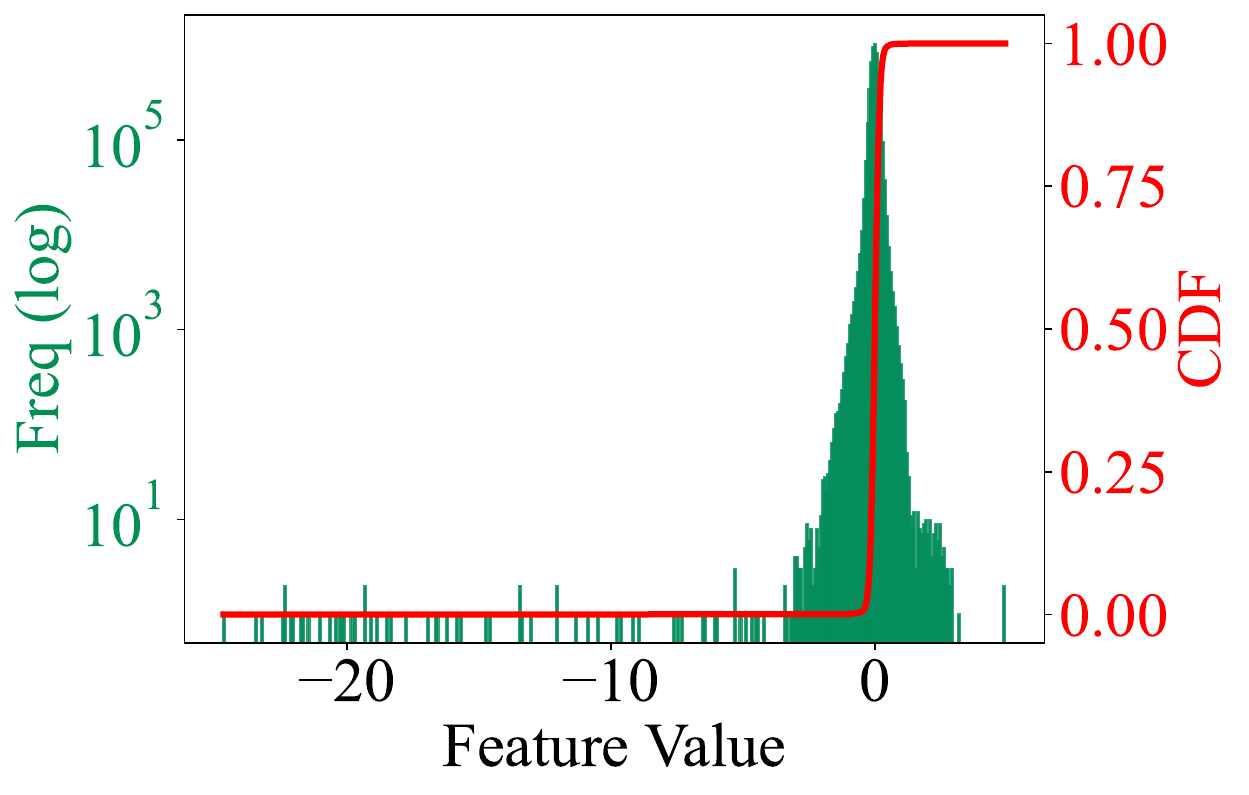}
	\end{minipage}
    \begin{minipage}{0.45\columnwidth}
		\centerline{\tiny{$p2$}}
		\vspace{0.2em}
		\includegraphics[width=\linewidth]{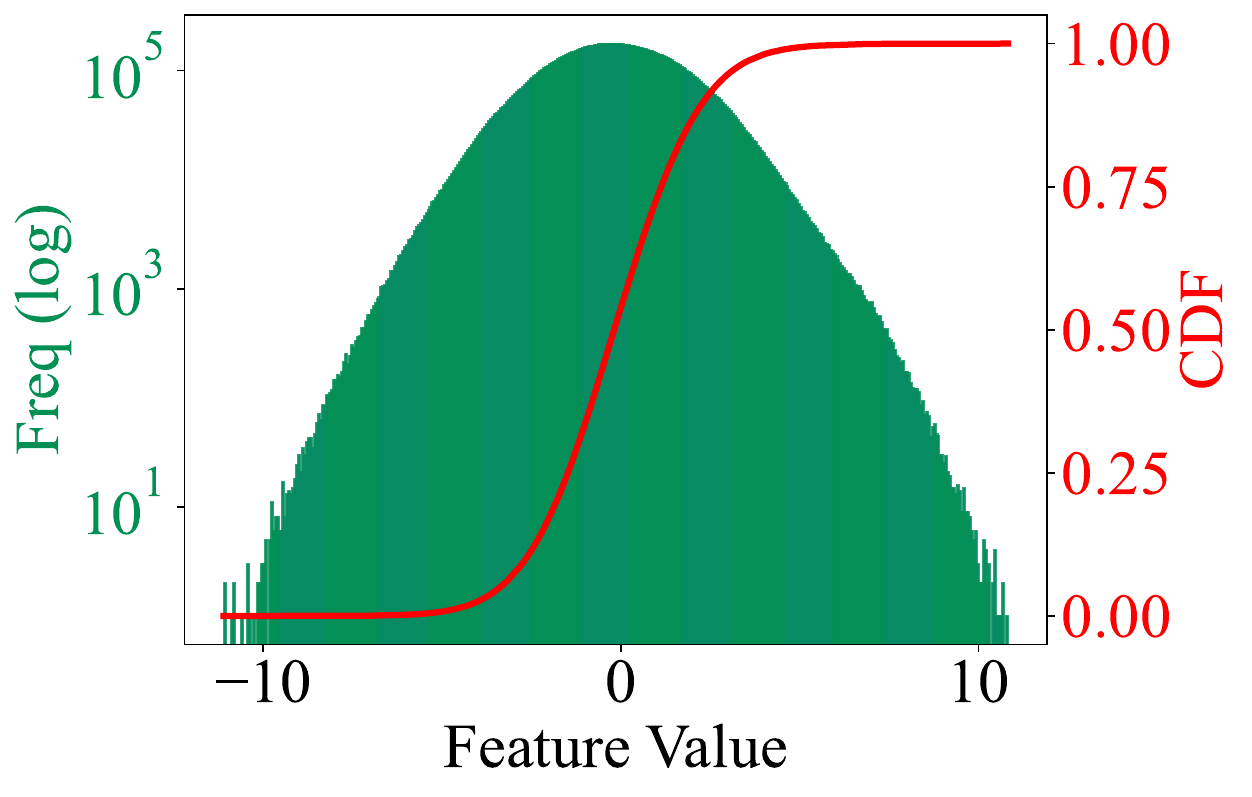}
	\end{minipage}
	\begin{minipage}{0.45\columnwidth}
		\centerline{\tiny{$p3$}}
		\vspace{0.2em}
		\includegraphics[width=\linewidth]{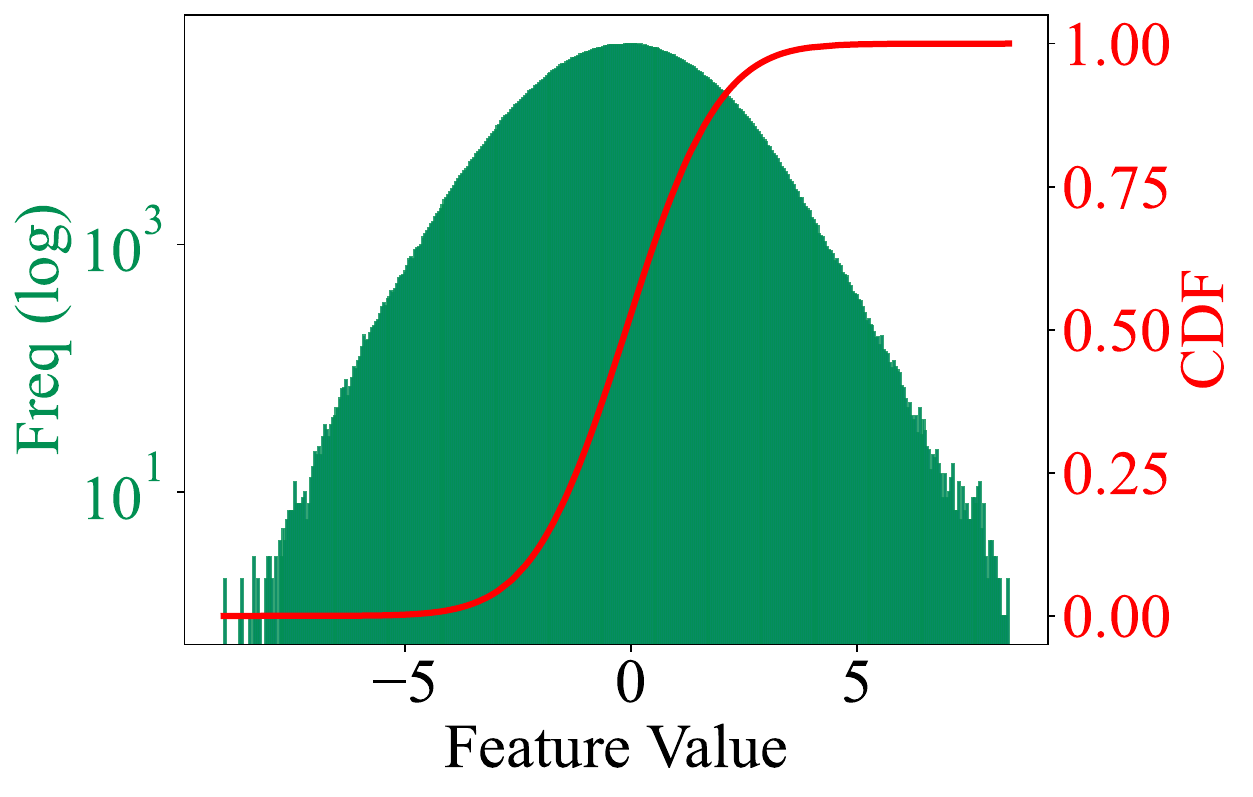}
	\end{minipage}\\
    
	\begin{minipage}{0.45\columnwidth}
		\centerline{\tiny{$SP_{DM3}$}}
		\vspace{0.2em}
		\includegraphics[width=\linewidth]{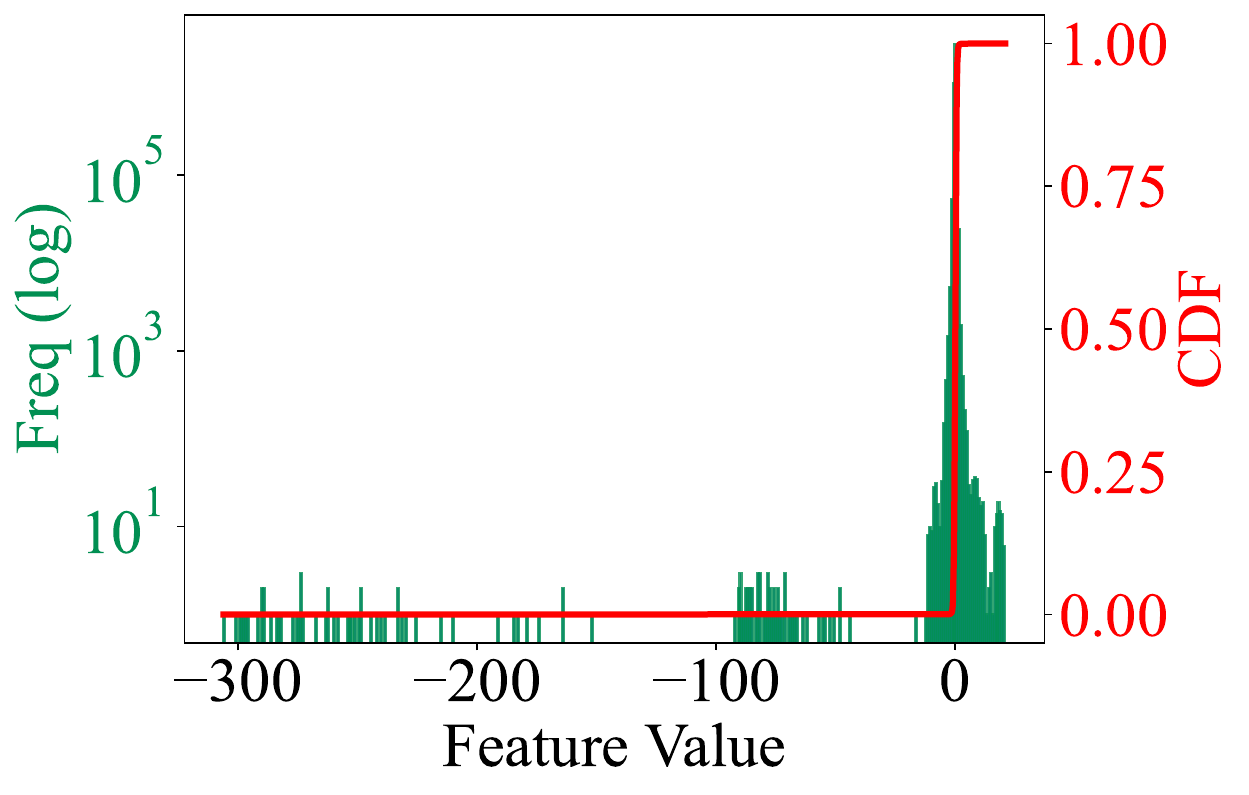}
	\end{minipage}
	\begin{minipage}{0.45\columnwidth}
		\centerline{\tiny{$SP_{DM4}$}}
		\vspace{0.2em}
		\includegraphics[width=\linewidth]{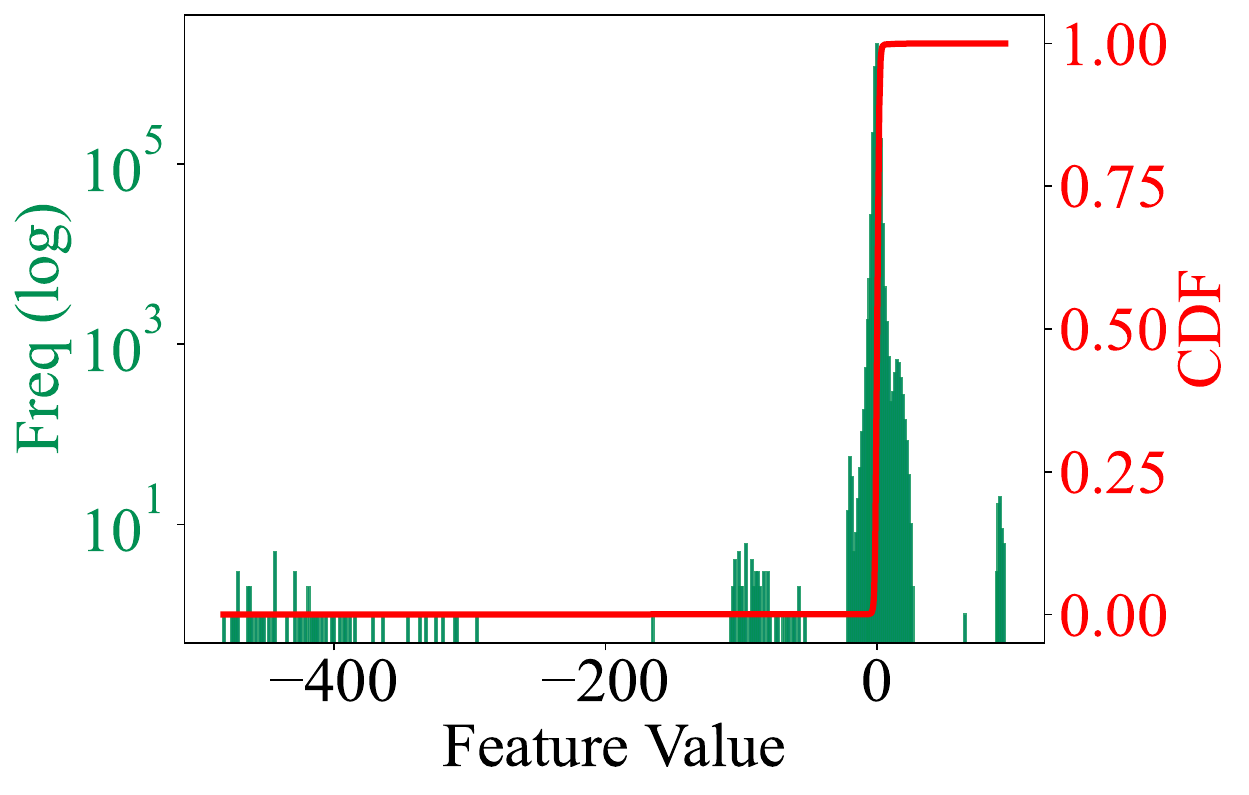}
	\end{minipage}
    \begin{minipage}{0.45\columnwidth}
		\centerline{\tiny{$p4$}}
		\vspace{0.2em}
		\includegraphics[width=\linewidth]{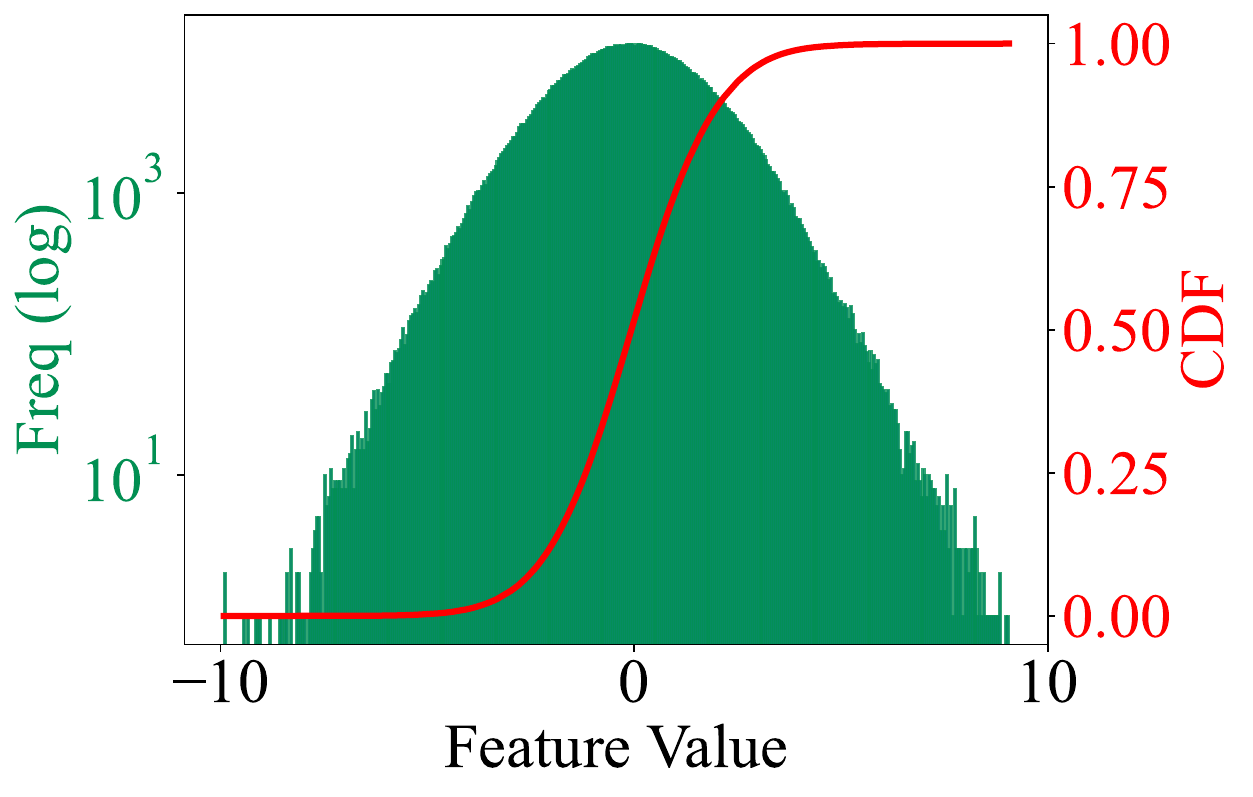}
	\end{minipage}
	\begin{minipage}{0.45\columnwidth}
		\centerline{\tiny{$p5$}}
		\vspace{0.2em}
		\includegraphics[width=\linewidth]{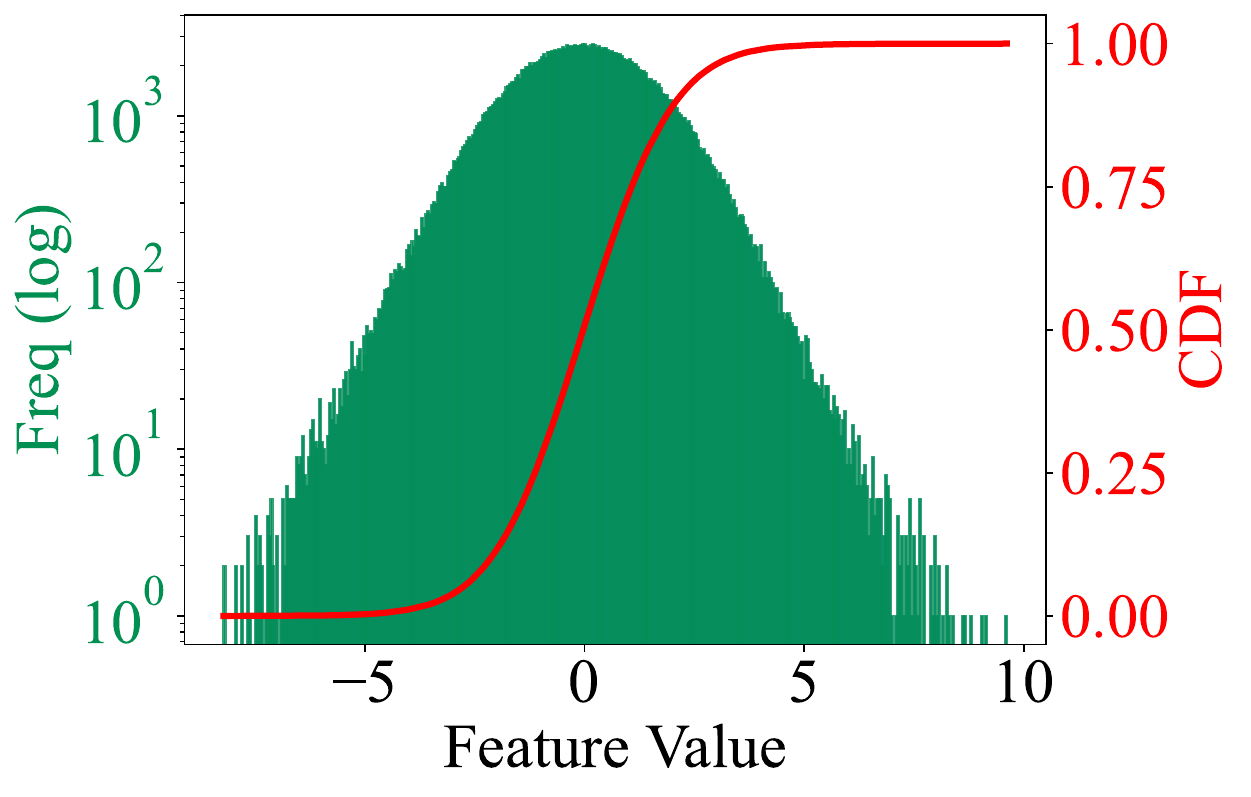}
	\end{minipage}\\
    
	\begin{minipage}{0.45\columnwidth}
		\centerline{\tiny{$SP_{GS}$}}
		\vspace{0.2em}
		\includegraphics[width=\linewidth]{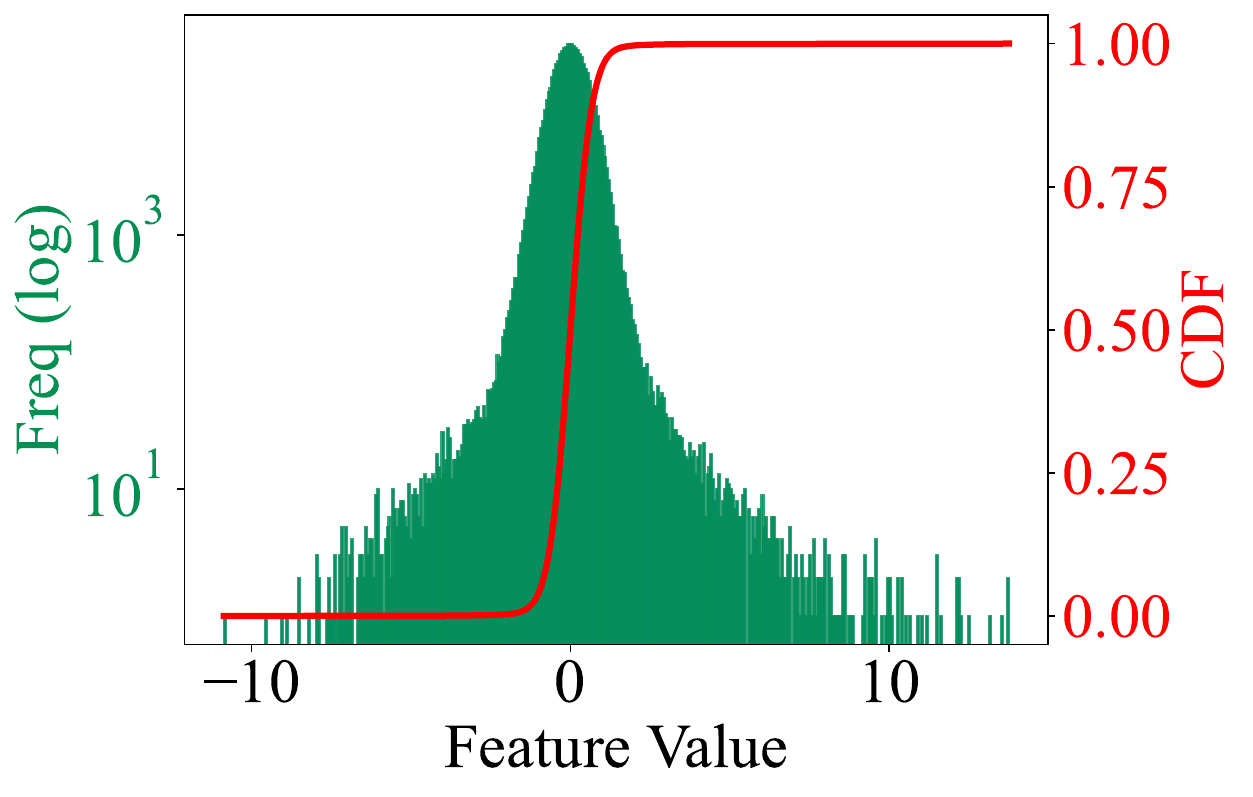}
	\end{minipage}
	\begin{minipage}{0.45\columnwidth}
		\centerline{\tiny{$SP_{H}$}}
		\vspace{0.2em}
		\includegraphics[width=\linewidth]{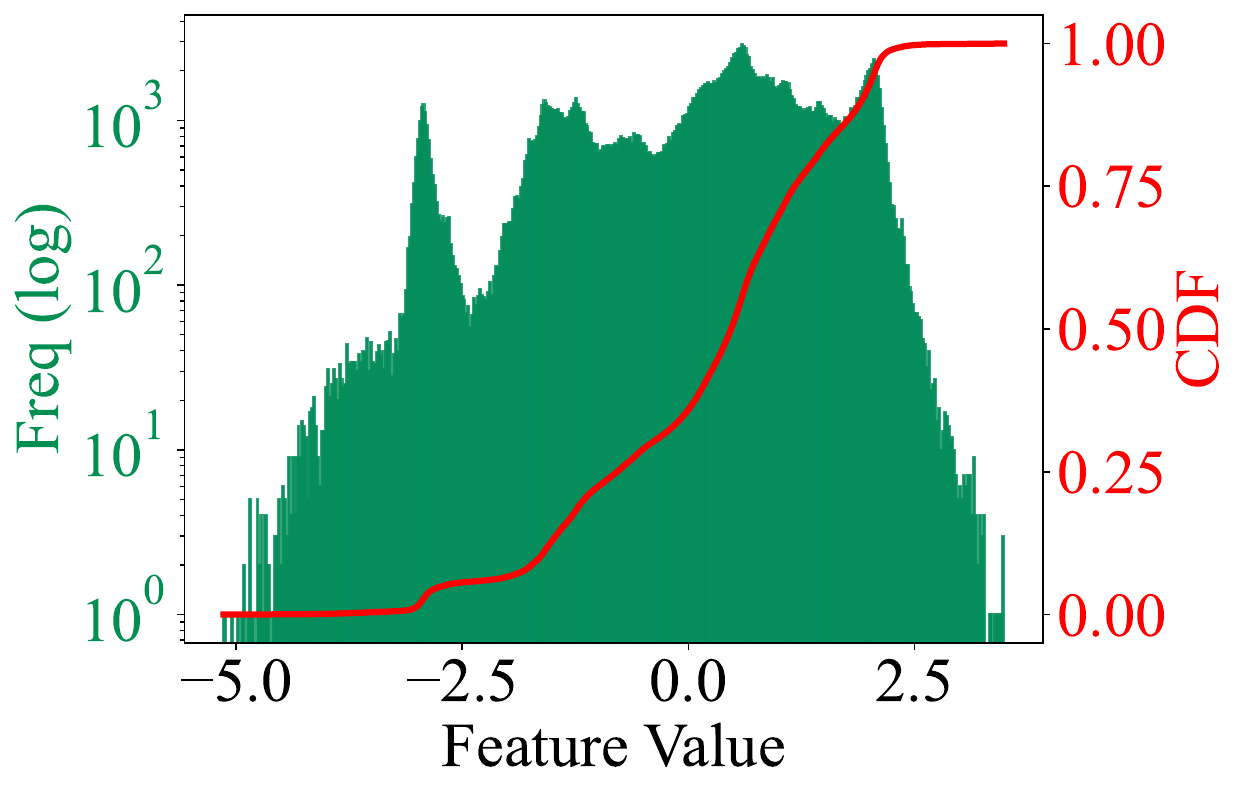}
	\end{minipage}
    \begin{minipage}{0.45\columnwidth}
		\centerline{\tiny{$C5$}}
		\vspace{0.2em}
		\includegraphics[width=\linewidth]{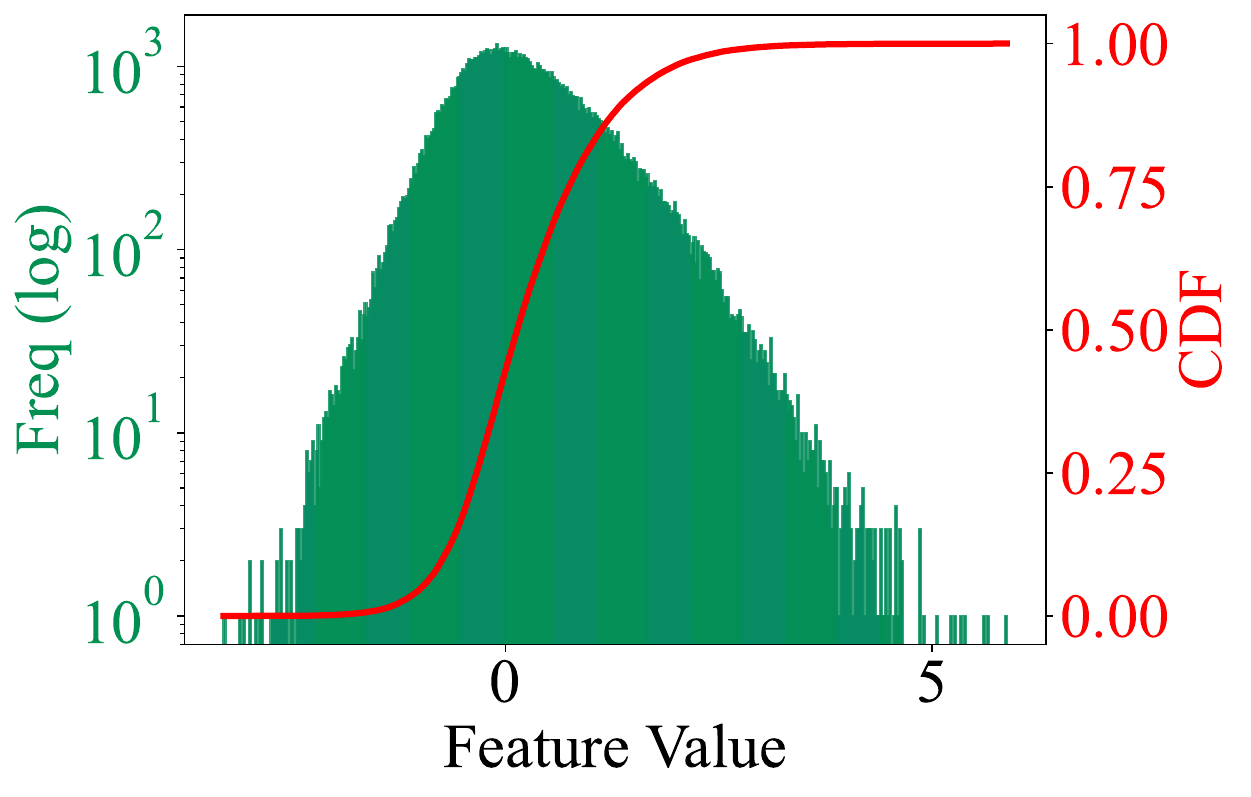}
	\end{minipage}
    \begin{minipage}{0.45\columnwidth}
		\centerline{\tiny{$C5$-ReLU}}
		\vspace{0.2em}
		\includegraphics[width=\linewidth]{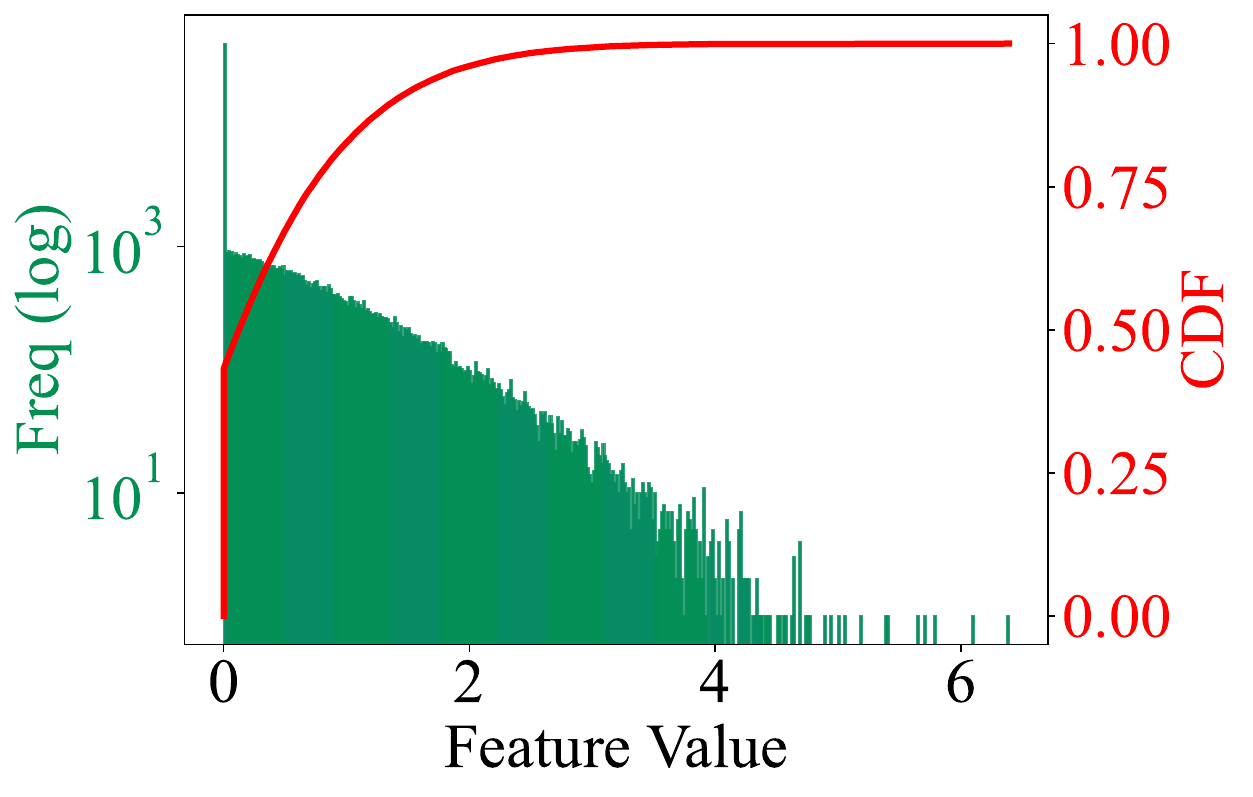}
	\end{minipage}%}
	\caption{Frequency and CDF comparisons between features in the proposed dataset and commonly used existing features. The proposed features exhibit distinct distributions compared to existing features, highlighting the necessity and value of the dataset. In addition, the proposed features demonstrate greater diversity, enhancing their representativeness and suitability for long-term use. Frequencies are scaled logarithmically for better visualization.}
	\label{fig_frequency_cdf}
\end{figure*}
\subsubsection{Cloud-edge distributed training}
\label{sec_cloud_edge_training}
In addition to computational resources, large model training requires access to large-scale datasets, which are often only available to a few major high-tech companies. Furthermore, sensitive data, such as medical records and financial information, cannot be freely shared due to privacy concerns and regulatory constraints.
To leverage distributed, privately owned data, split learning has emerged as a viable solution \cite{vepakomma2018split}. Split learning divides the training process into two phases, with the bulk of the computational load handled by the cloud, while edge devices, which are often resource-constrained, handle a smaller portion of the task. This approach minimizes the computational demand on the edge while enabling collaboration with powerful cloud resources.

The cloud-edge distributed training pipeline is depicted in the middle of  Fig. \ref{fig_scenario}. The whole pipeline is divided into two parts: edge models and cloud models. During forward propagation, the edge model processes private data to generate features, which are then encoded into bitstreams and transmitted to the cloud. Upon receiving the bitstreams, the cloud decodes them back into features and feeds them into the cloud model. In the backward pass, gradients are computed and backpropagated along the reverse path. 
% To optimize this process (reduce bandwidth costs and minimize latency), it is critical to apply feature coding before transmission.
To reduce bandwidth cost and minimize latency, it is critical to apply feature coding before transmission.

\subsection{Cloud-edge distributed inference}
Once large models are trained, they are deployed in products to provide client services. For a product or service to succeed, two primary concerns must be addressed: supporting a high volume of simultaneous cloud server access and protecting user privacy.
For the first concern, managing computational load is essential, particularly in generative applications that involve repetitive diffusion processes. A practical solution is to offload part of this computational demand to edge devices. For the second concern, the same strategy in the cloud-edge distributed training, as discussed in Sec. \ref{sec_cloud_edge_training}, can be applied.

The cloud-edge distributed inference pipeline is illustrated on the right of Fig. \ref{fig_scenario}. In the upload phase, source data is converted into features by an edge model on the client side. These features are then encoded and transmitted to the cloud, where they are further processed by large models. In the download phase, the processed features are sent back to the client to complete a specific task. Depending on the task head used, various functionalities can be performed with the returned features. For example, Stable Diffusion 3 (SD3) features derived from an image can be used to synthesize video or generate segmentation masks.
This approach not only reduces the computational load on the cloud server but also enhances data privacy by keeping raw user data on the edge side.
\begin{figure*}[tbp]
    \centering
    % first row
    \vspace{-4mm}
    \begin{subfigure}[t]{0.135\linewidth}
        \centering
        \caption*{\tiny{$SP_{DM1}$}}
        \includegraphics[width=\linewidth]{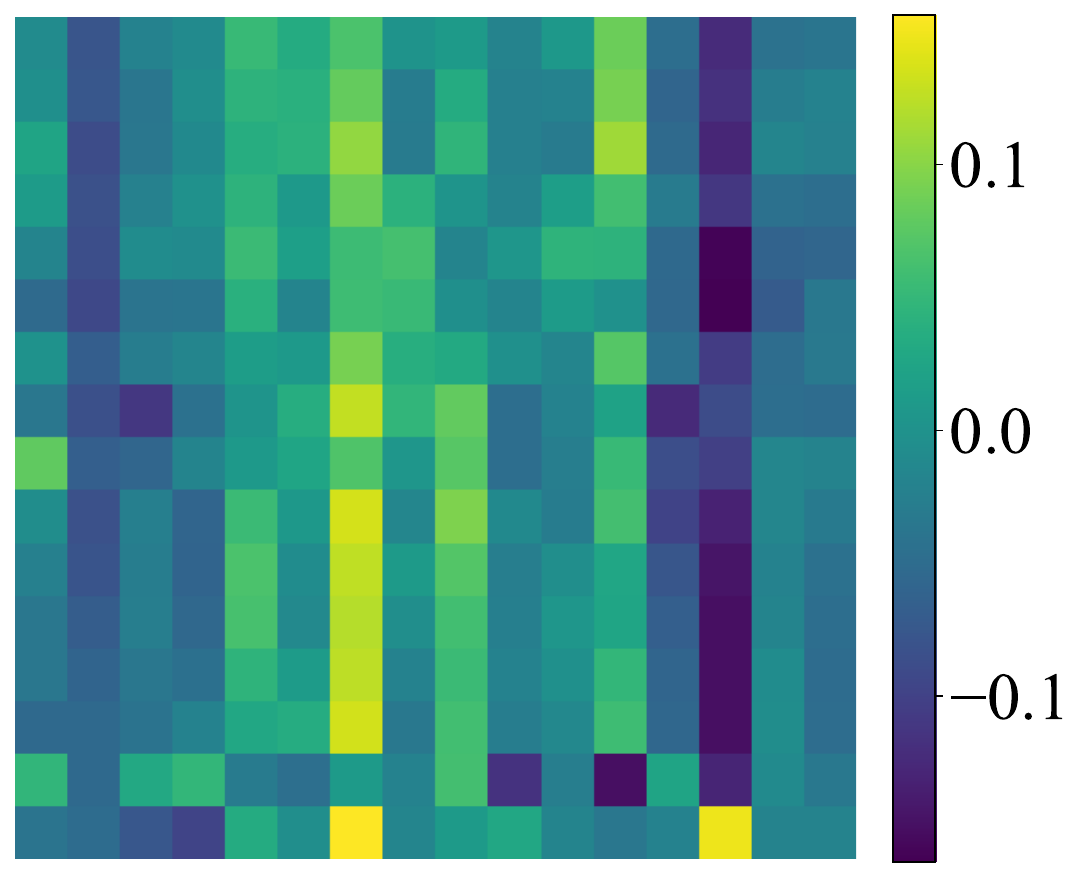}
    \end{subfigure}
    \begin{subfigure}[t]{0.135\linewidth}
        \centering
        \caption*{\tiny{$SP_{DM2}$}}
        \includegraphics[width=\linewidth]{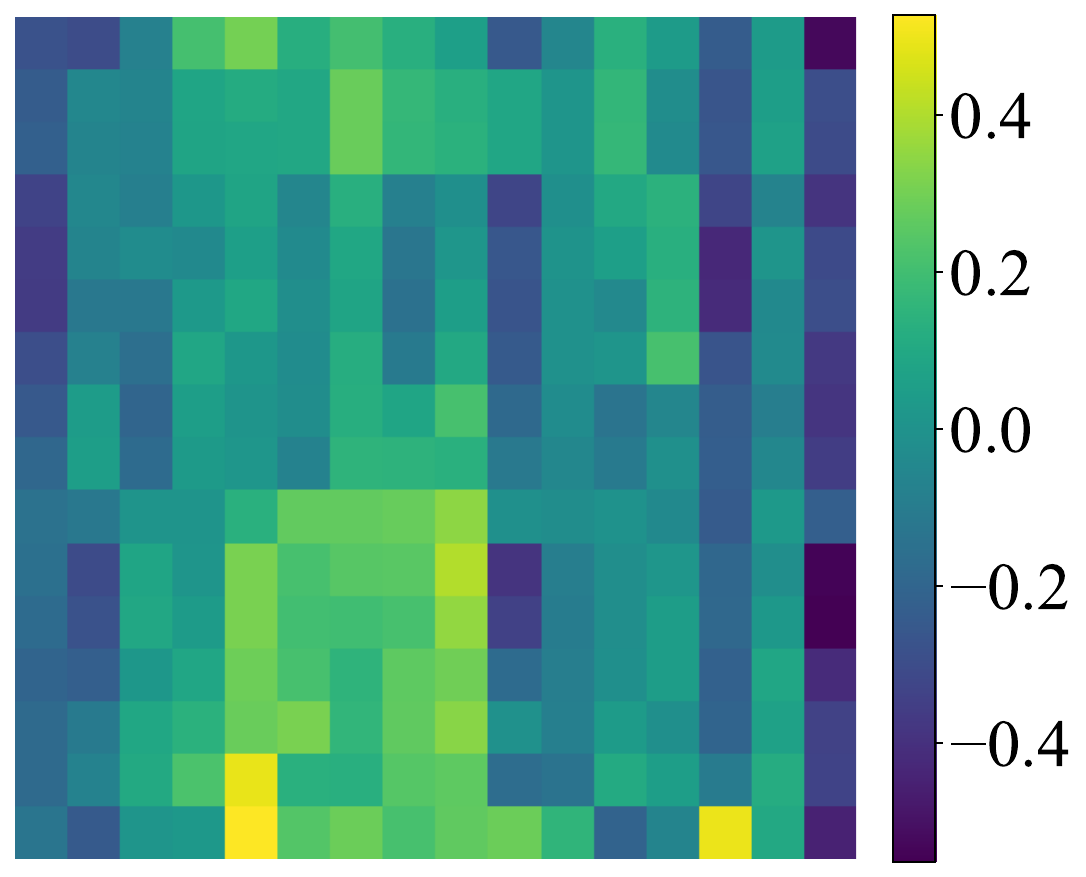}
    \end{subfigure}
    \begin{subfigure}[t]{0.135\linewidth}
        \centering
        \caption*{\tiny{$SP_{DM3}$}}
        \includegraphics[width=\linewidth]{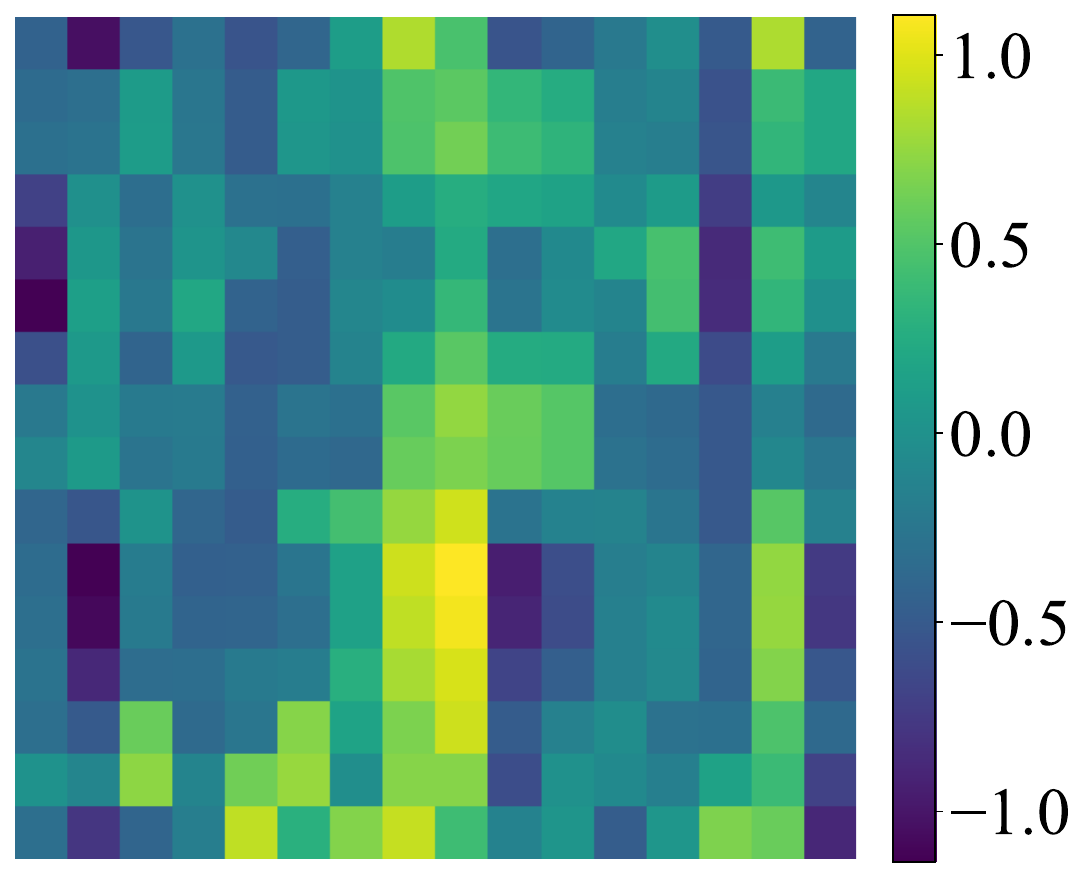}
    \end{subfigure}
    \begin{subfigure}[t]{0.1296\linewidth}
        \centering
        \caption*{\tiny{$SP_{DM4}$}}
        \includegraphics[width=\linewidth]{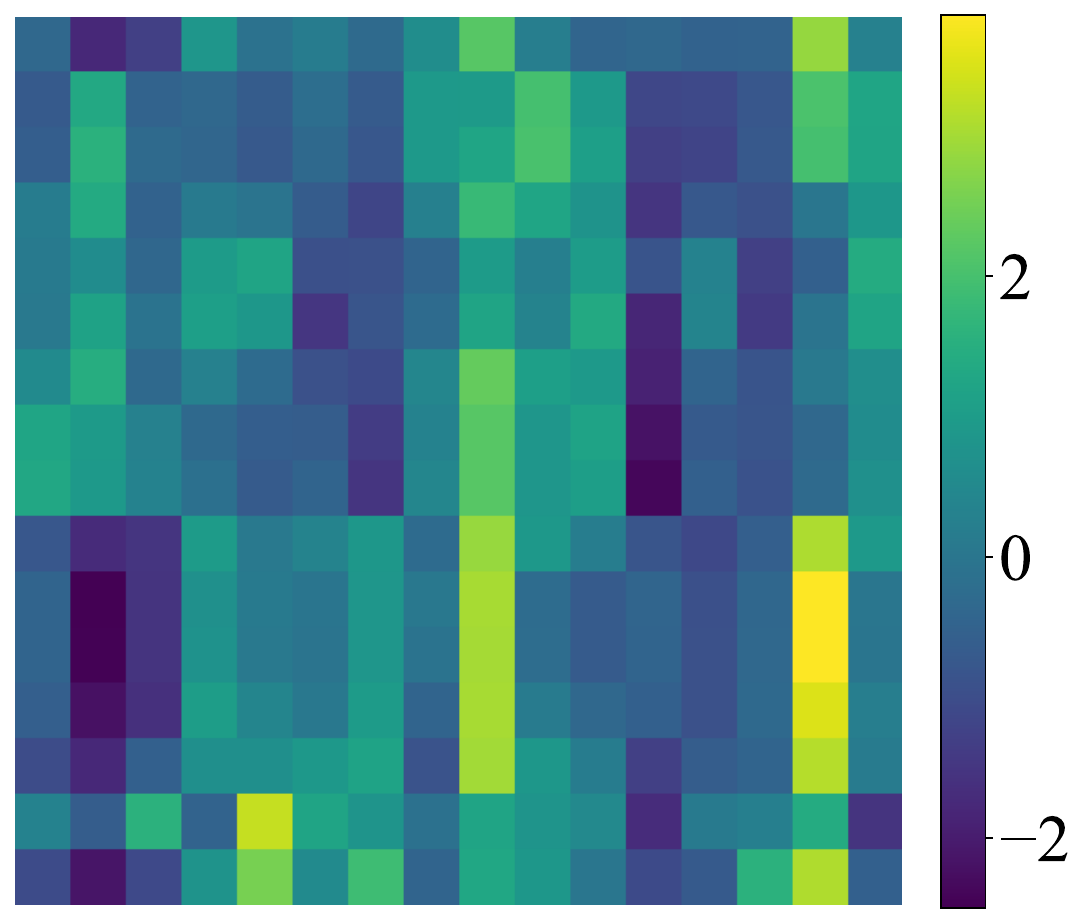}
    \end{subfigure}
    \begin{subfigure}[t]{0.1296\linewidth}
        \centering
        \caption*{\tiny{$SP_{GS}$}}
        \includegraphics[width=\linewidth]{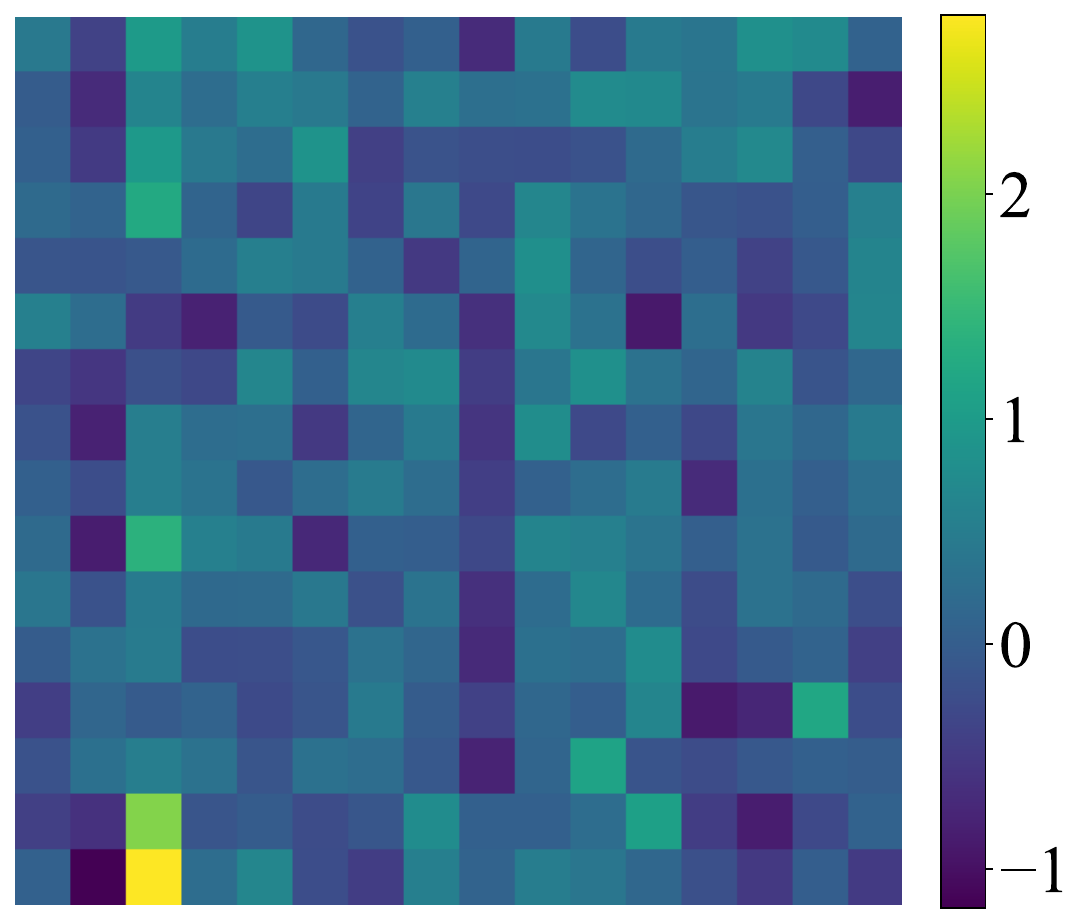}
    \end{subfigure}
    \begin{subfigure}[t]{0.1314\linewidth}
        \centering
        \caption*{\tiny{$SP_{H}$}}
        \includegraphics[width=\linewidth]{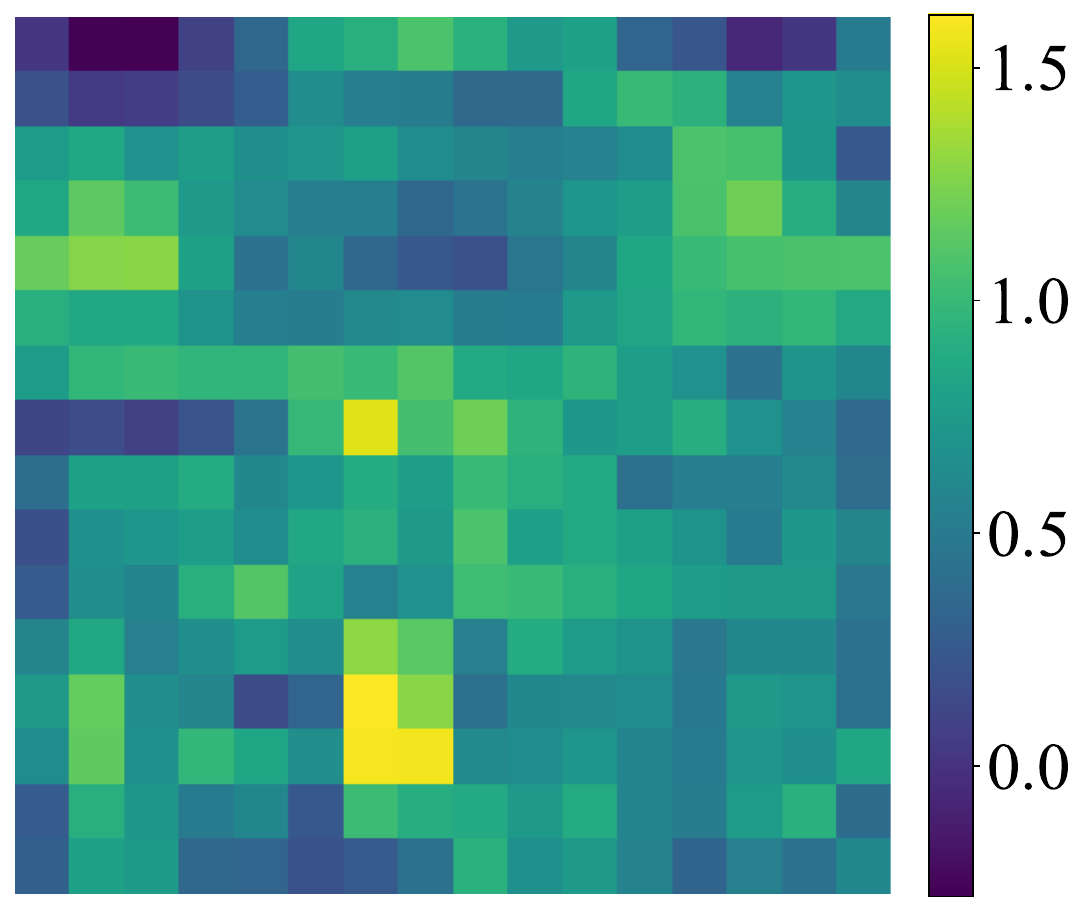}
    \end{subfigure}
    \vspace{0.2mm}

    % second row
    \begin{subfigure}[t]{0.135\linewidth}
        \centering
        \includegraphics[width=\linewidth]{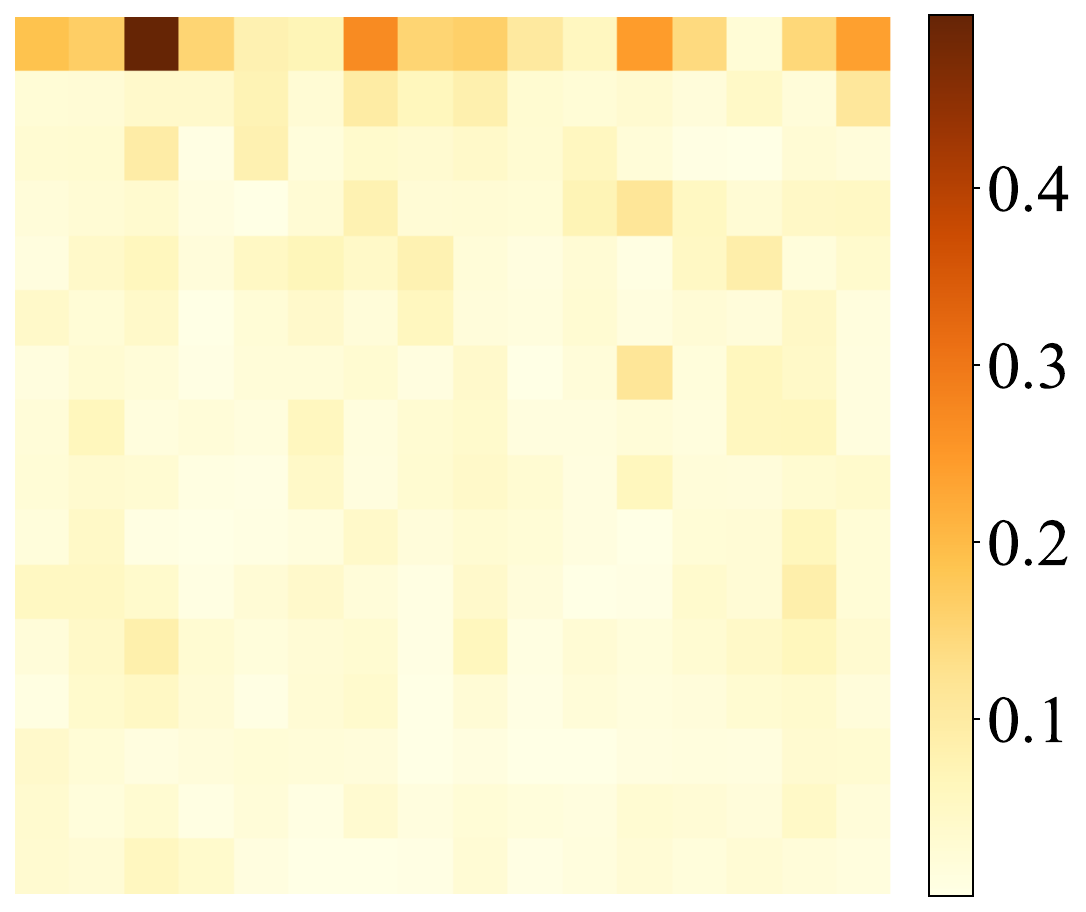}
    \end{subfigure}
    \begin{subfigure}[t]{0.135\linewidth}
        \centering
        \includegraphics[width=\linewidth]{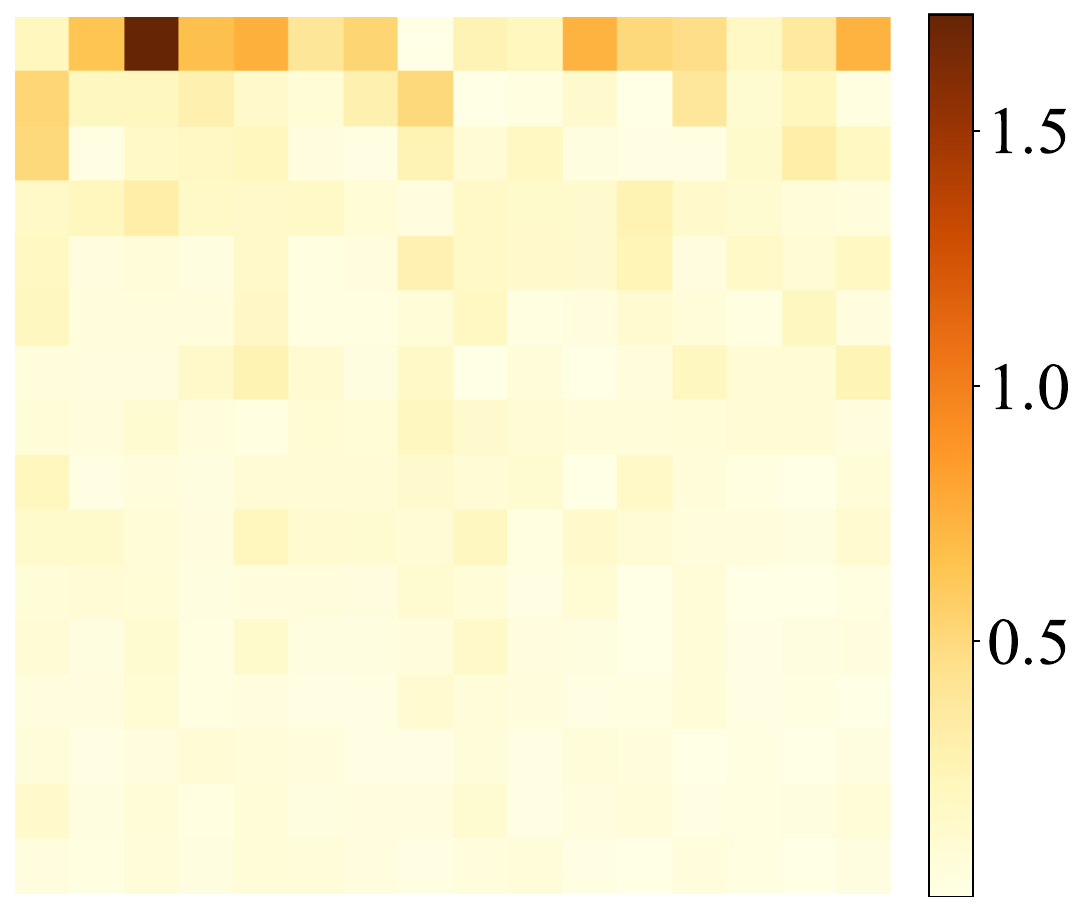}
    \end{subfigure}
    \begin{subfigure}[t]{0.135\linewidth}
        \centering
        \includegraphics[width=\linewidth]{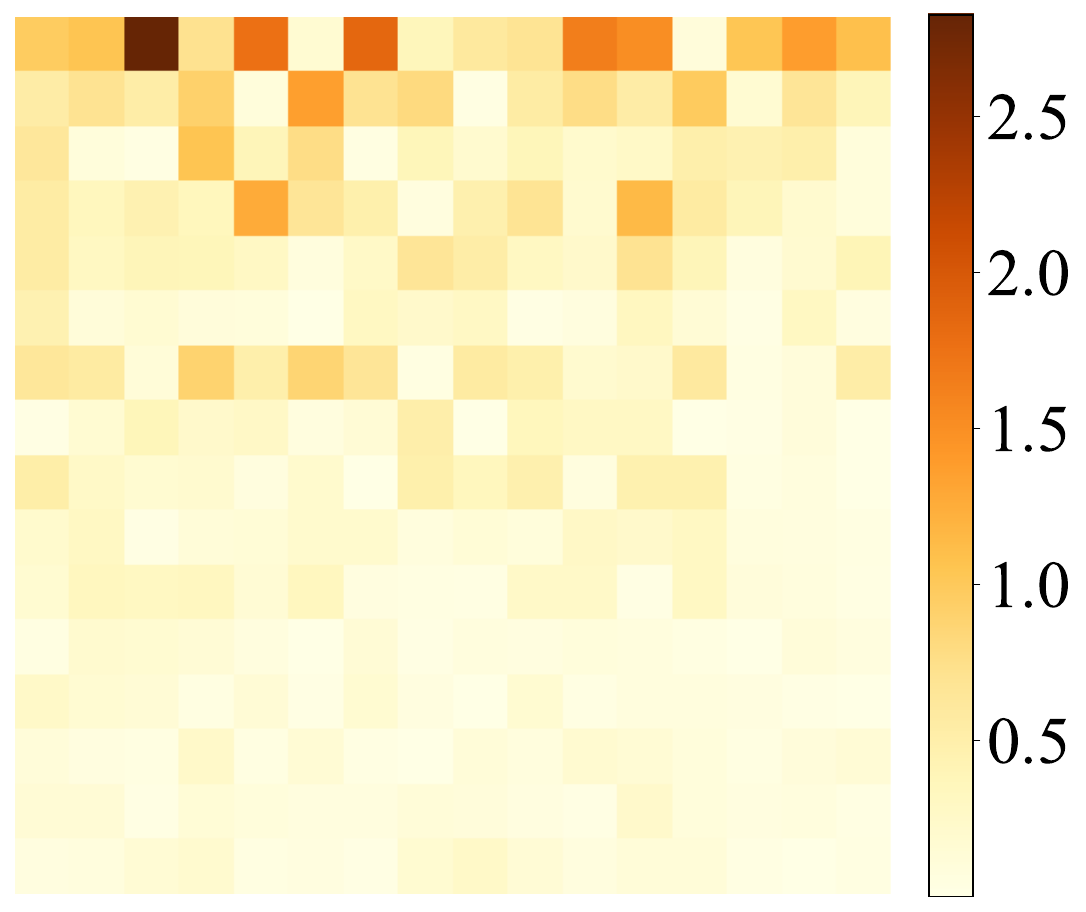}
    \end{subfigure}
    \begin{subfigure}[t]{0.1287\linewidth}
        \centering
        \includegraphics[width=\linewidth]{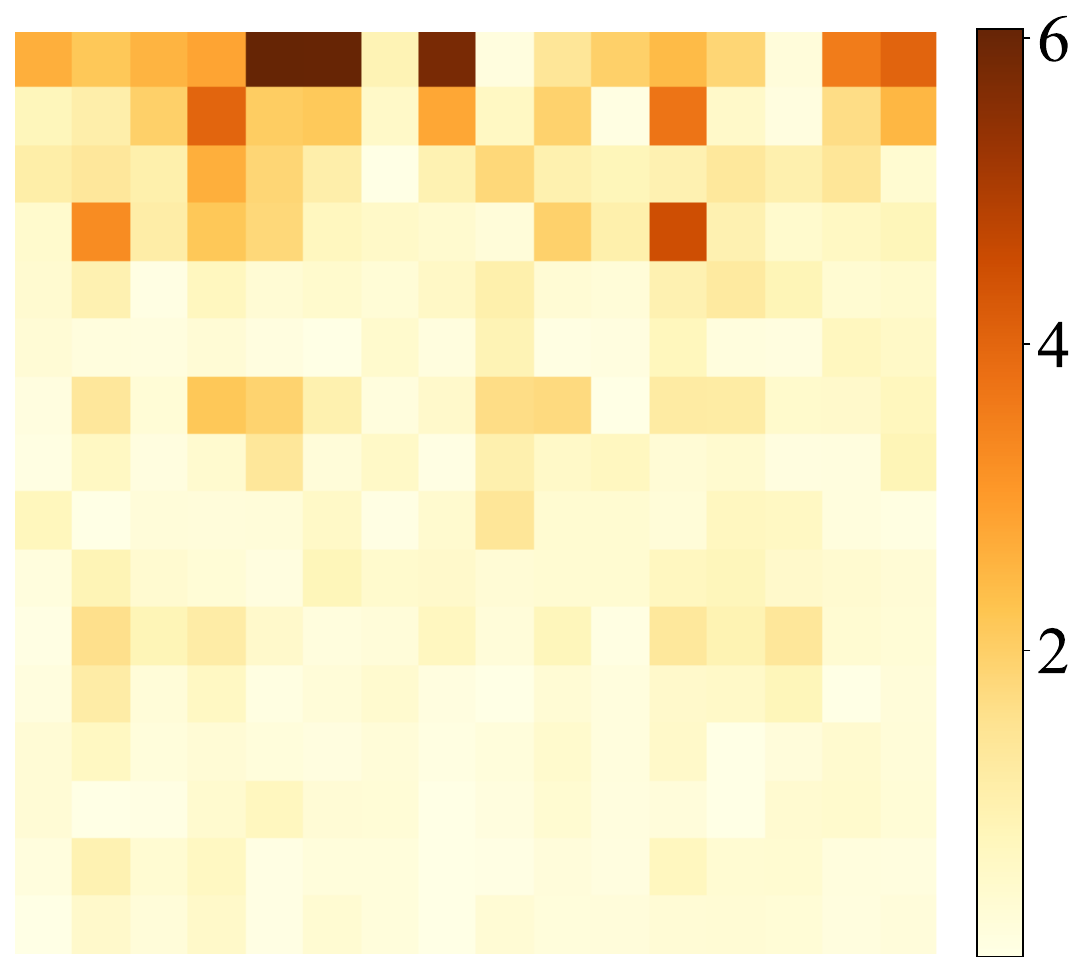}
    \end{subfigure}
    \begin{subfigure}[t]{0.135\linewidth}
        \centering
        \includegraphics[width=\linewidth]{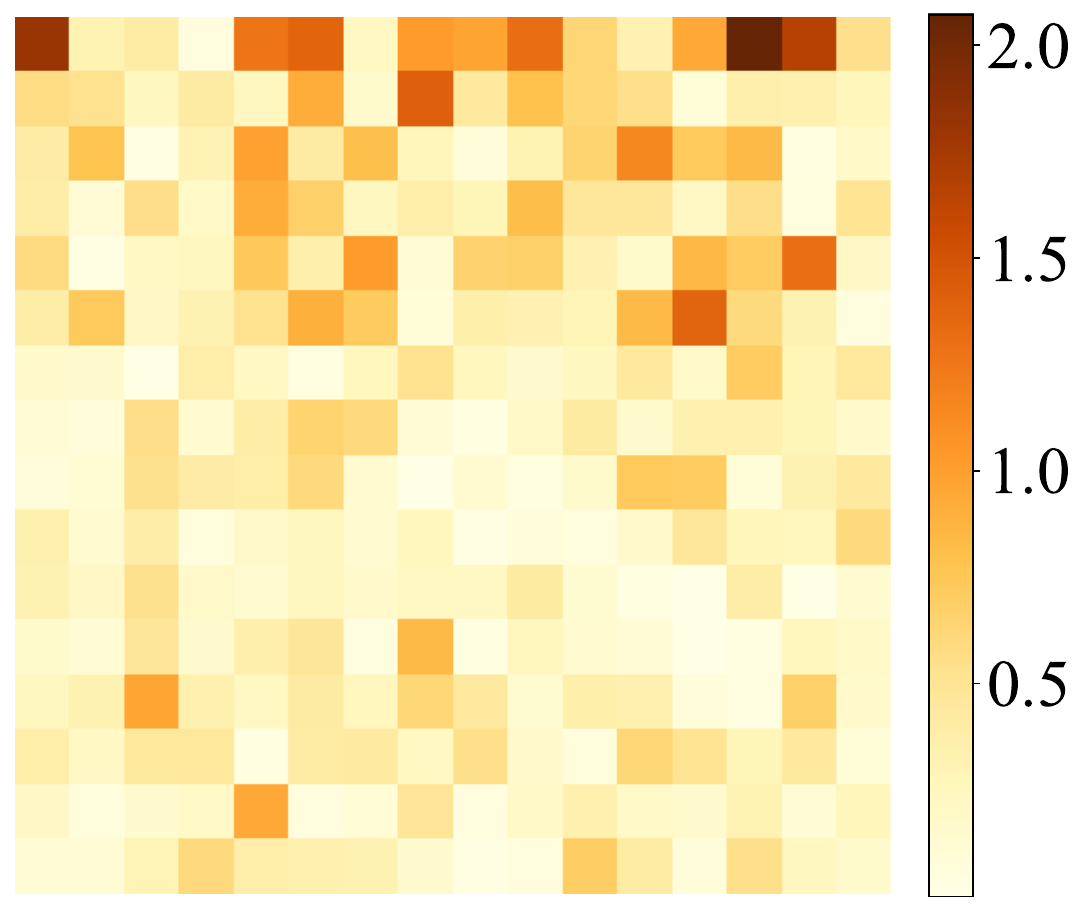}
    \end{subfigure}
    \begin{subfigure}[t]{0.135\linewidth}
        \centering
        \includegraphics[width=\linewidth]{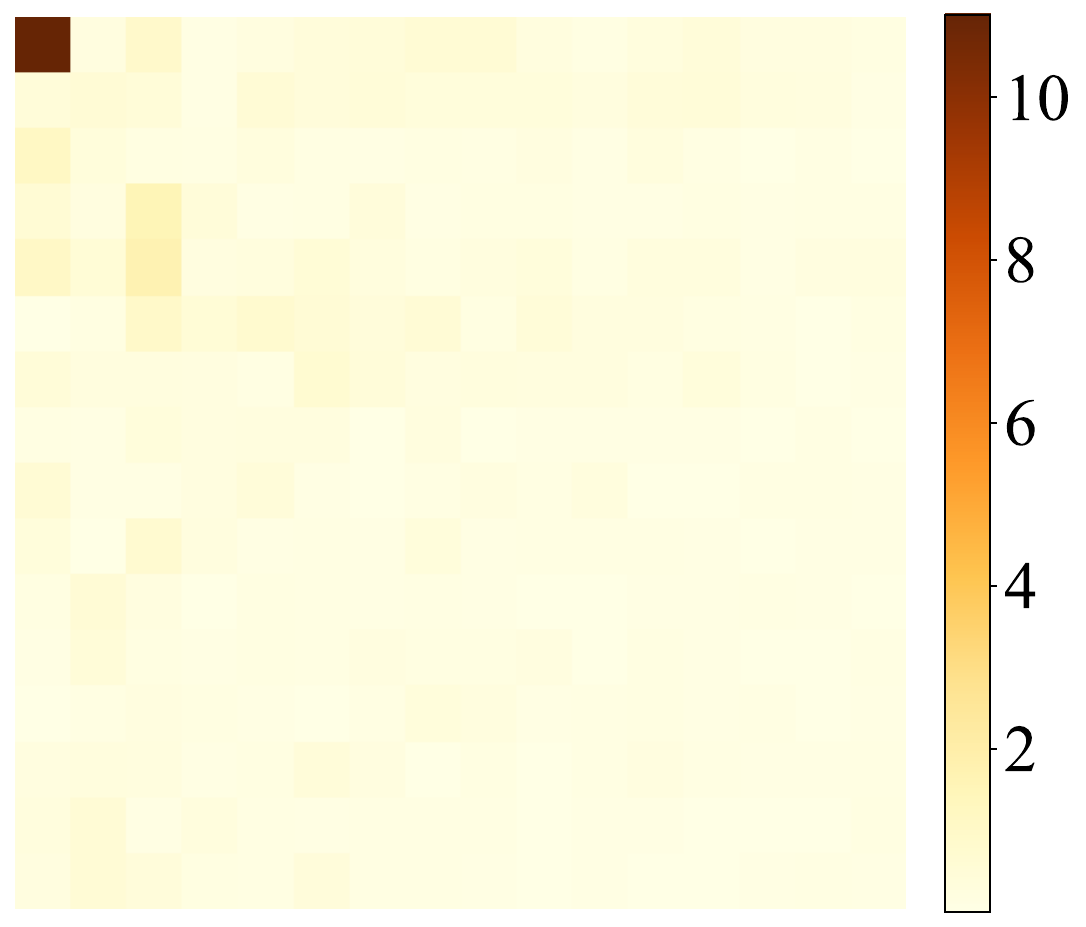}
    \end{subfigure}

    % third row
    \vspace{-4mm}
    \begin{subfigure}[t]{0.1332\linewidth}
        \centering
        \caption*{\scriptsize{$p2$}}
        \includegraphics[width=\linewidth]{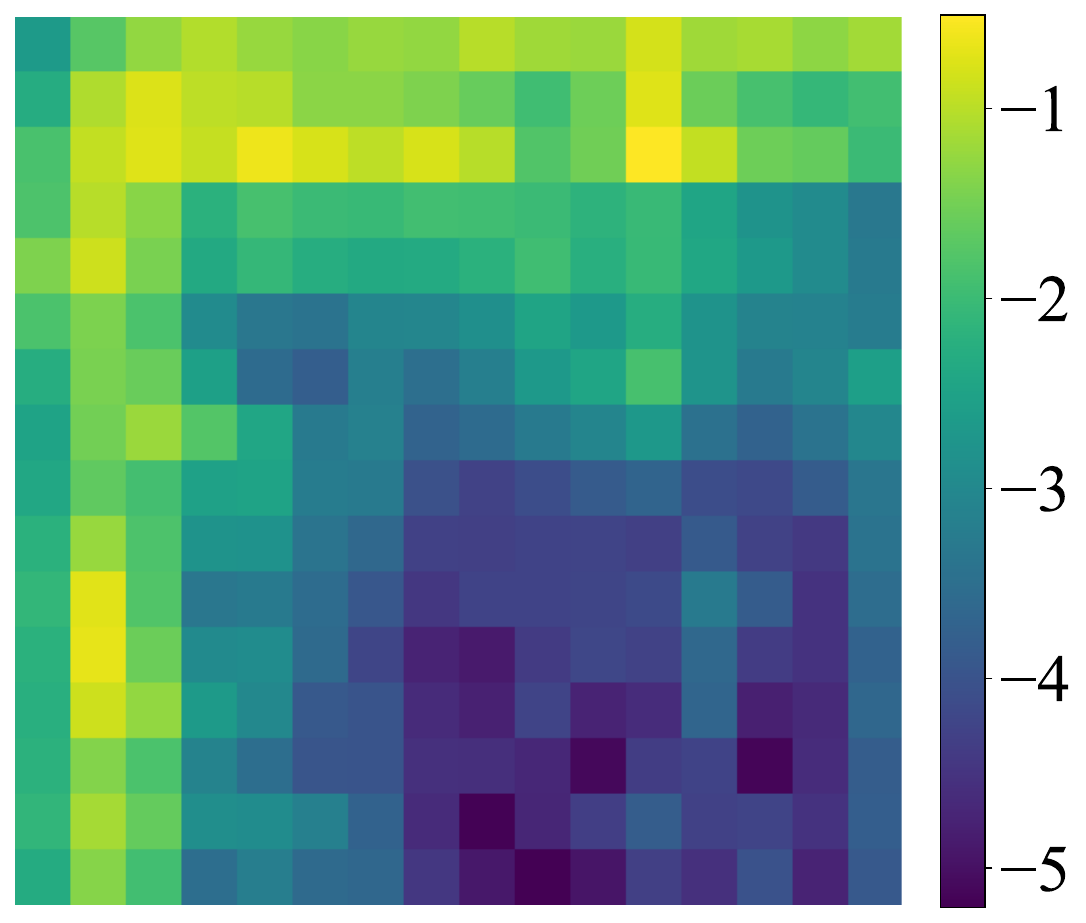}
    \end{subfigure}
    \begin{subfigure}[t]{0.1332\linewidth}
        \centering
        \caption*{\scriptsize{$p3$}}
        \includegraphics[width=\linewidth]{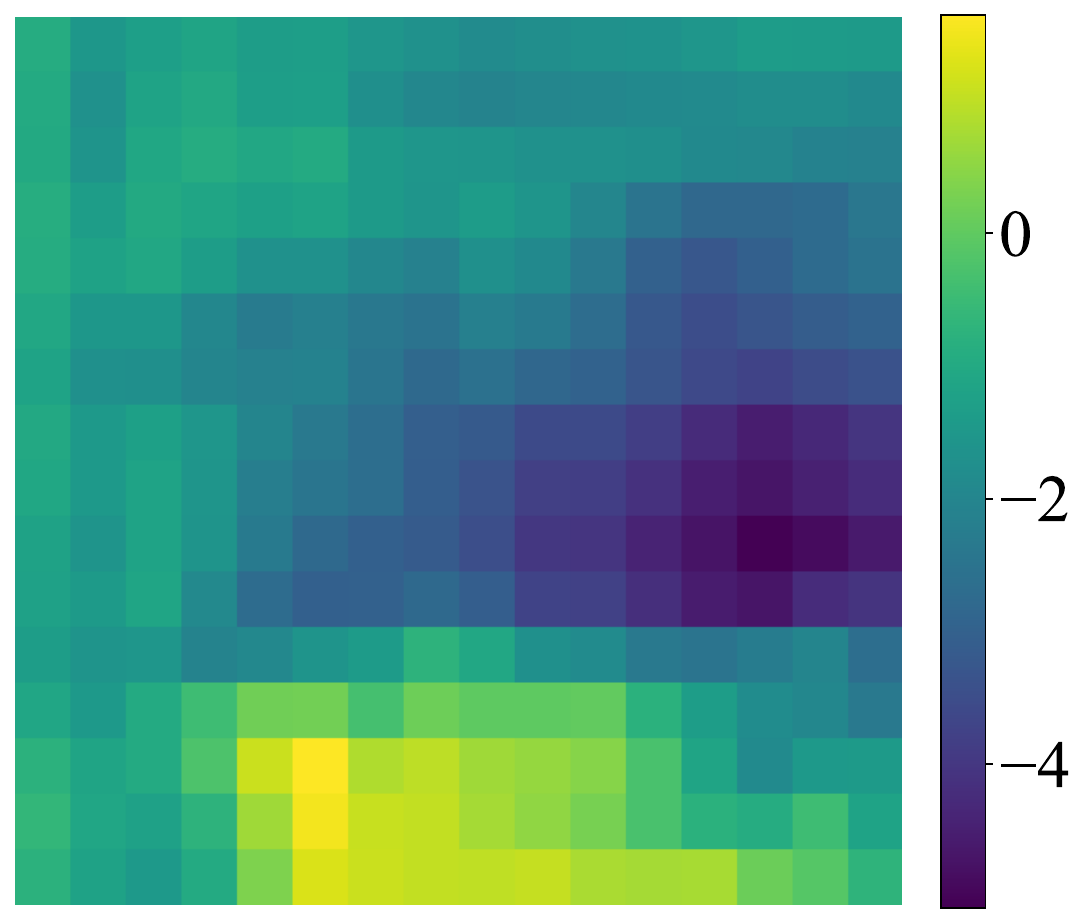}
    \end{subfigure}
    \begin{subfigure}[t]{0.1332\linewidth}
        \centering
        \caption*{\scriptsize{$p4$}}
        \includegraphics[width=\linewidth]{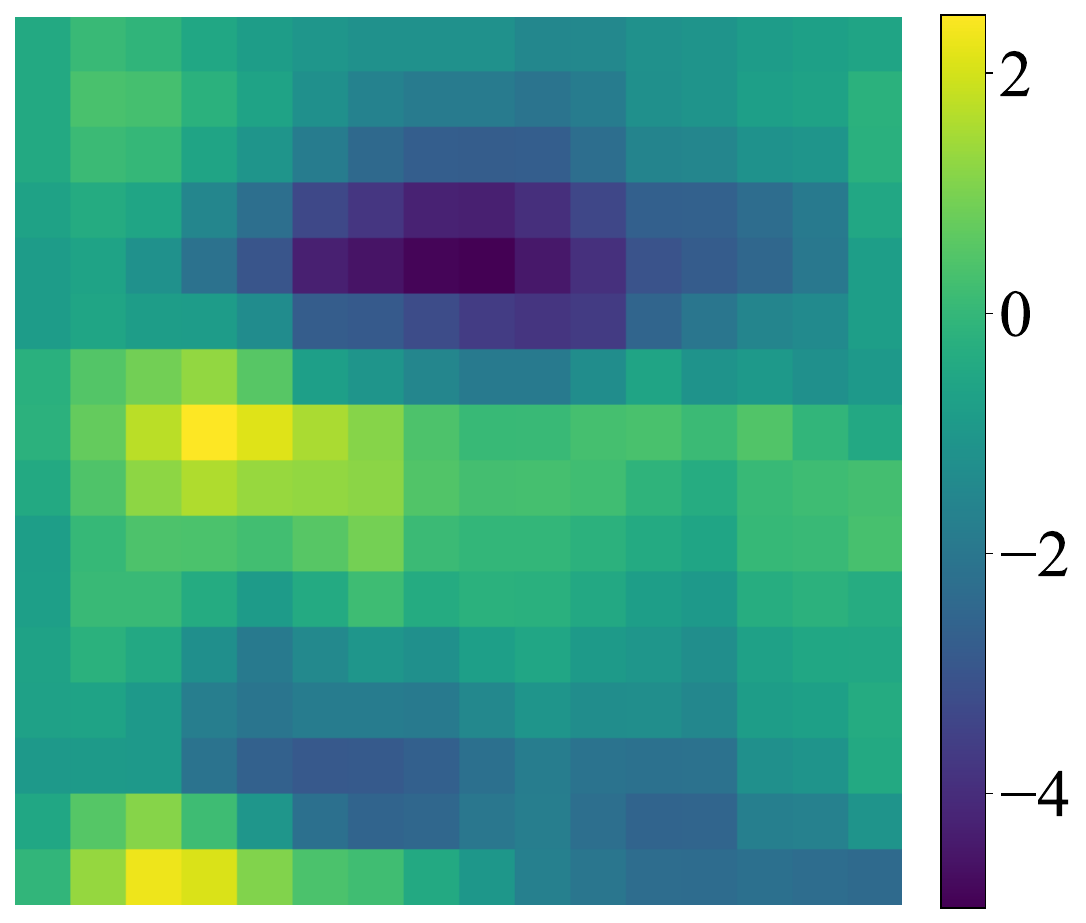}
    \end{subfigure}
    \begin{subfigure}[t]{0.1332\linewidth}
        \centering
        \caption*{\scriptsize{$p5$}}
        \includegraphics[width=\linewidth]{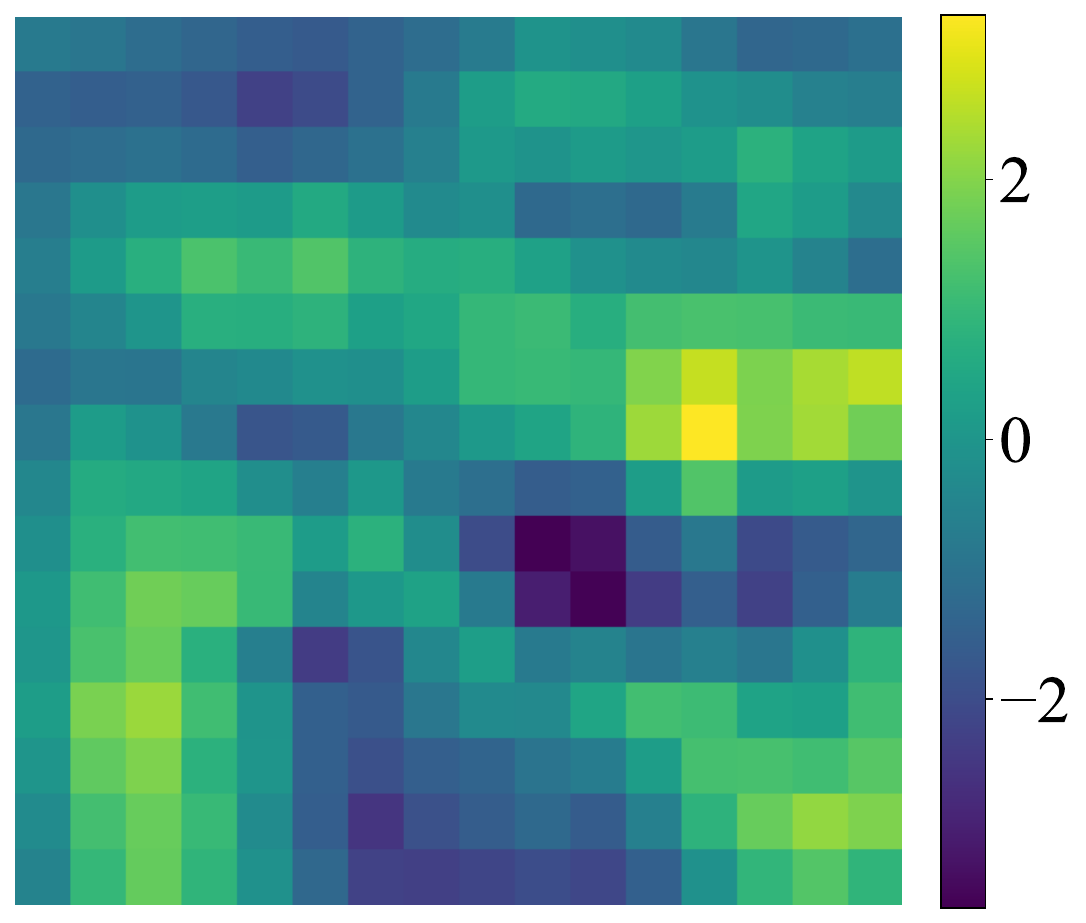}
    \end{subfigure}
    \begin{subfigure}[t]{0.1395\linewidth}
        \centering
        \caption*{\scriptsize{$C5$}}
        \includegraphics[width=\linewidth]{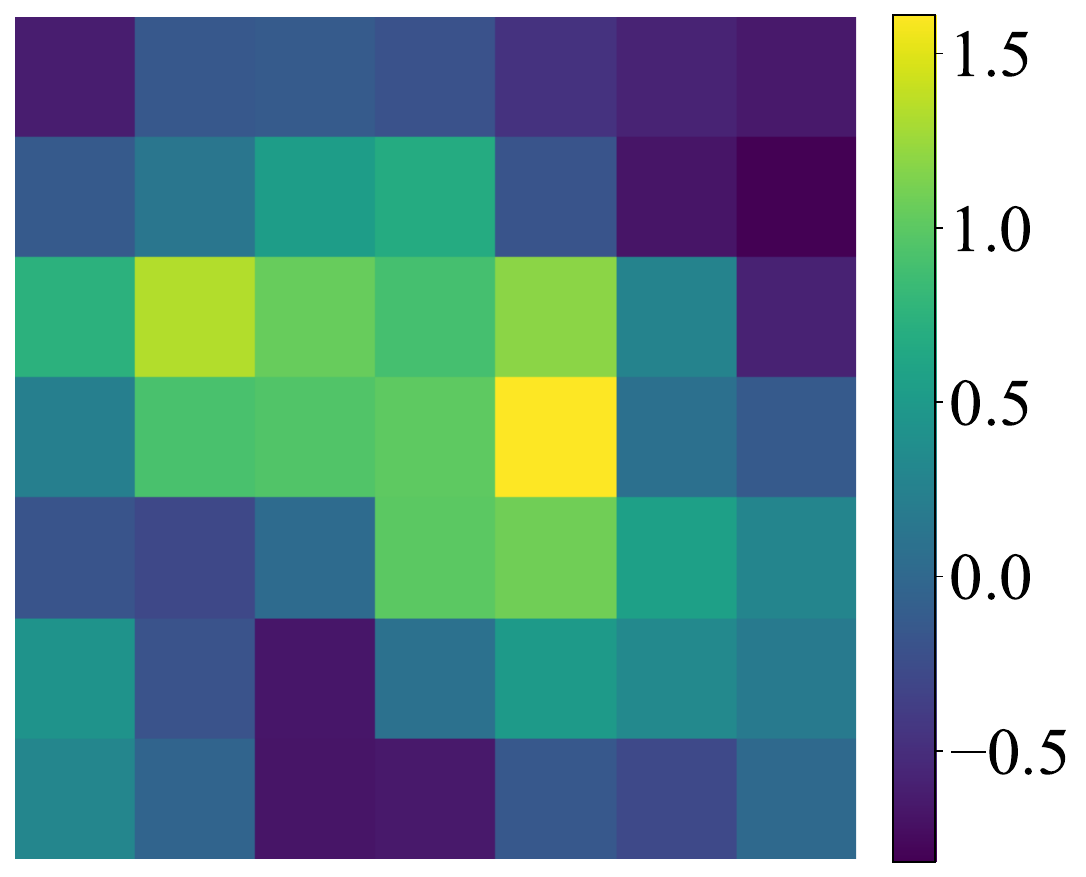}
    \end{subfigure}
    \begin{subfigure}[t]{0.135\linewidth}
        \centering
        \caption*{\scriptsize{$C5$-ReLU}}
        \includegraphics[width=\linewidth]{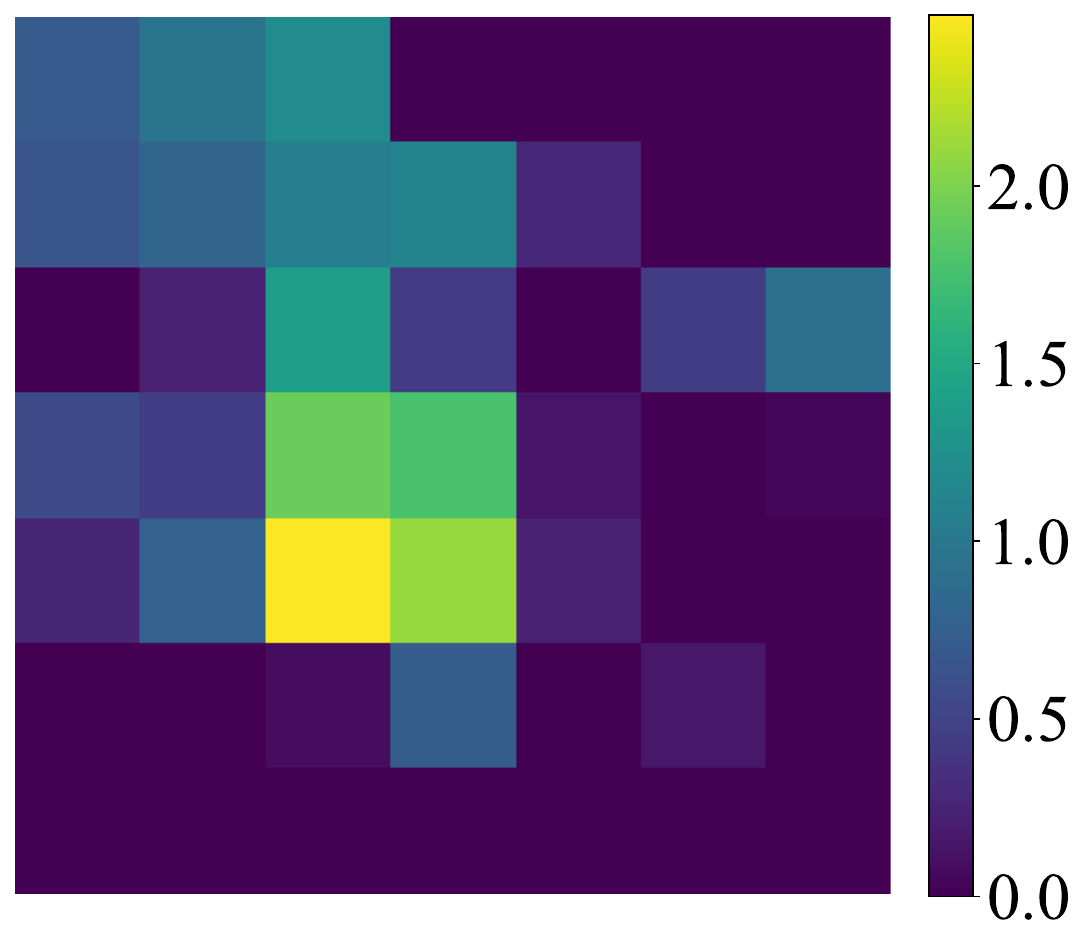}
    \end{subfigure}
    \vspace{0.2mm}

    % fourth row
    \begin{subfigure}[t]{0.135\linewidth}
        \centering
        \includegraphics[width=\linewidth]{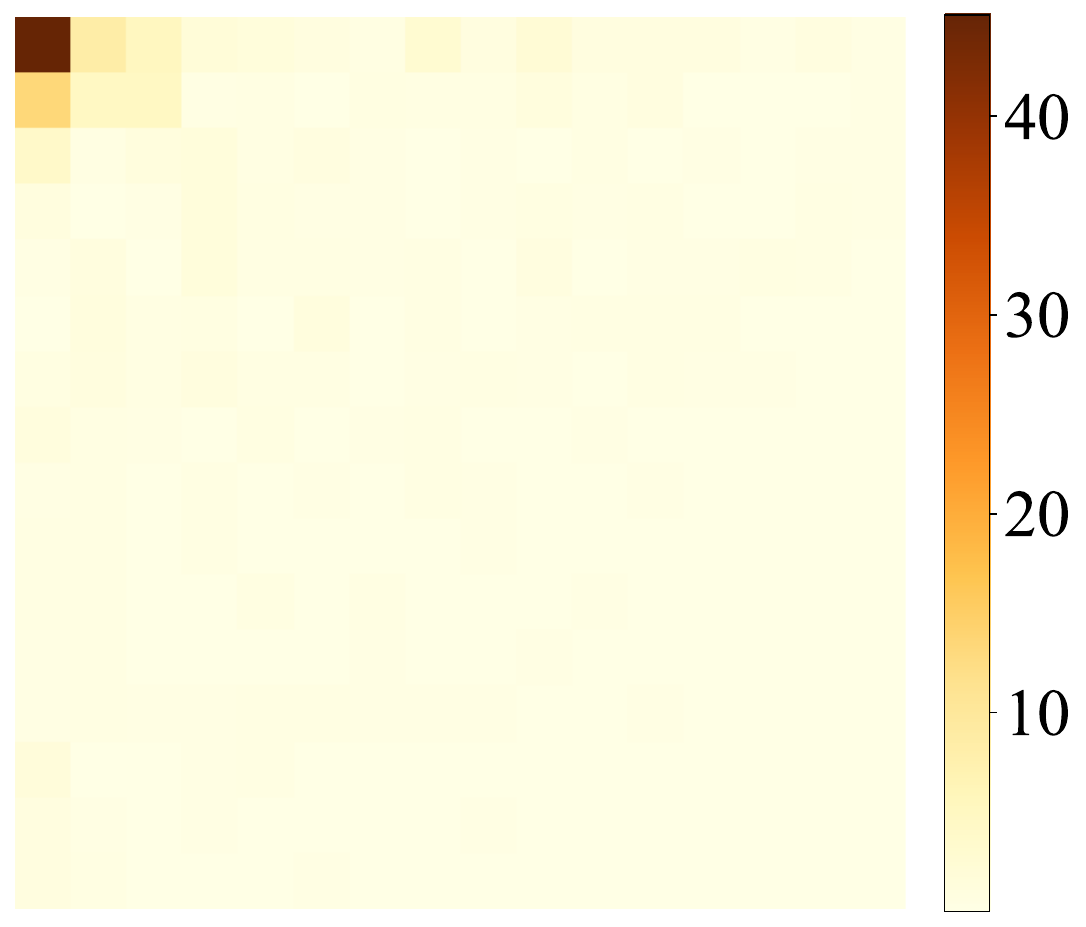}
    \end{subfigure}
    \begin{subfigure}[t]{0.135\linewidth}
        \centering
        \includegraphics[width=\linewidth]{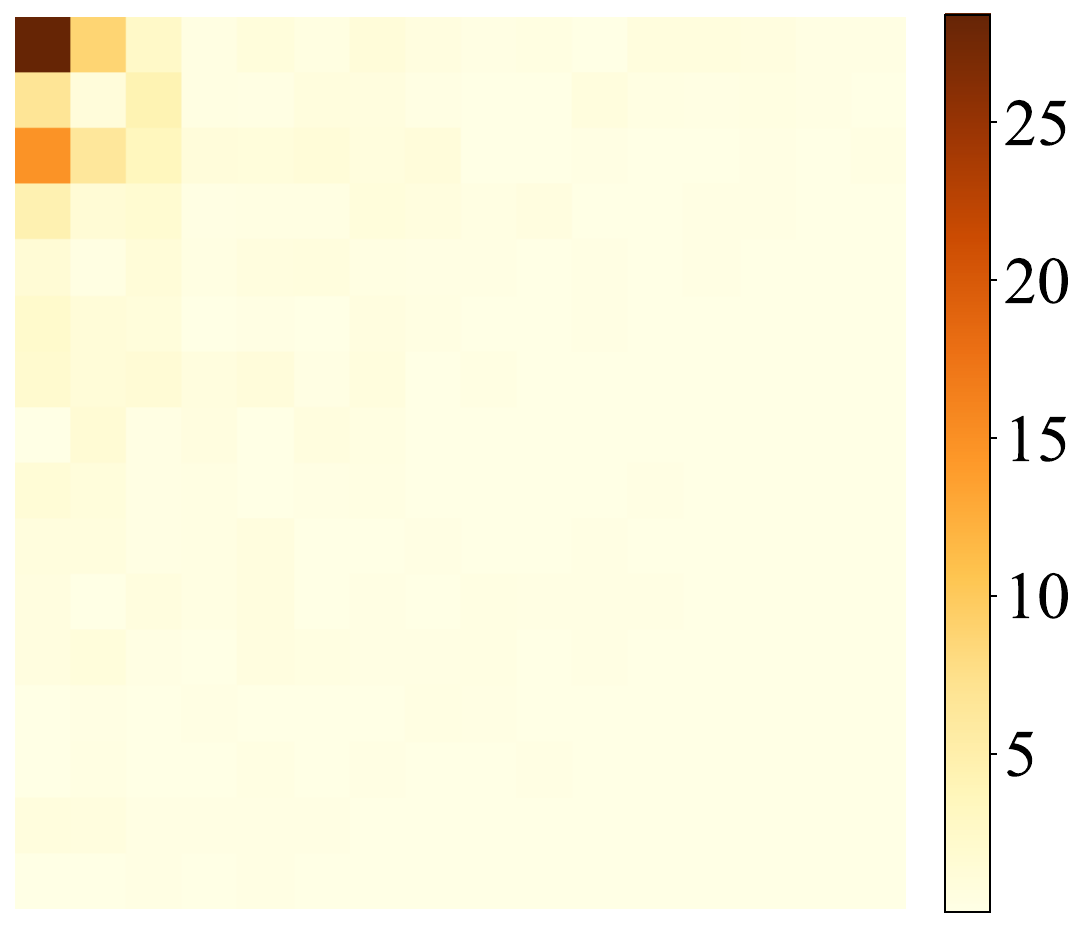}
    \end{subfigure}
    \begin{subfigure}[t]{0.135\linewidth}
        \centering
        \includegraphics[width=\linewidth]{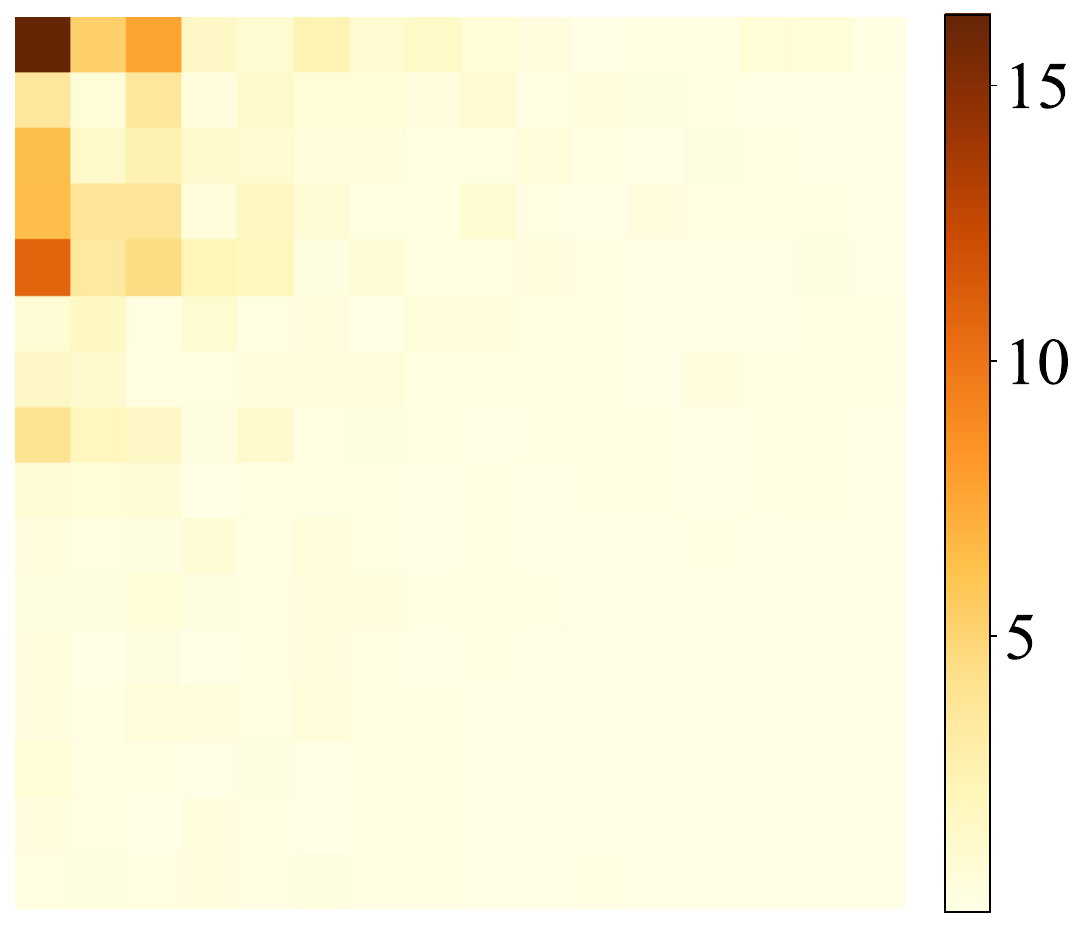}
    \end{subfigure}
    \begin{subfigure}[t]{0.1305\linewidth}
        \centering
        \includegraphics[width=\linewidth]{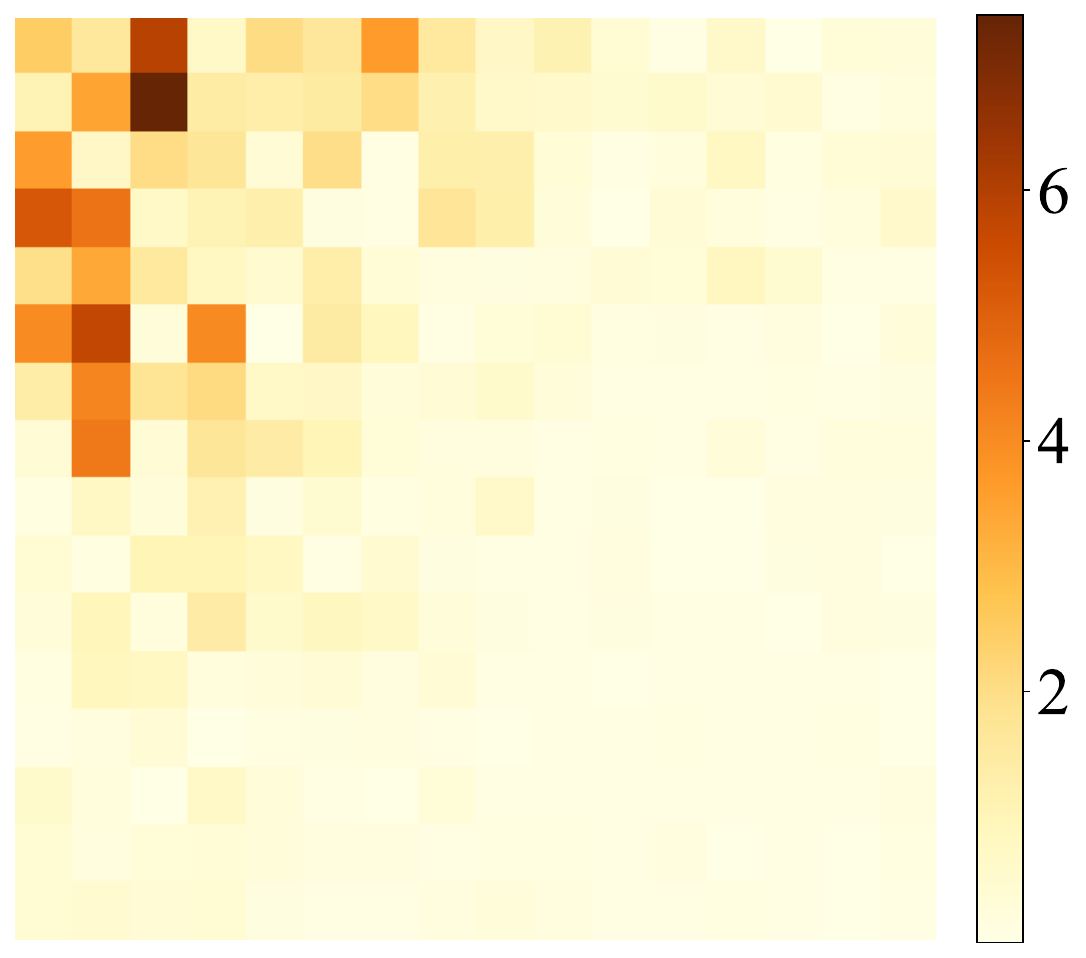}
    \end{subfigure}
    \begin{subfigure}[t]{0.1386\linewidth}
        \centering
        \includegraphics[width=\linewidth]{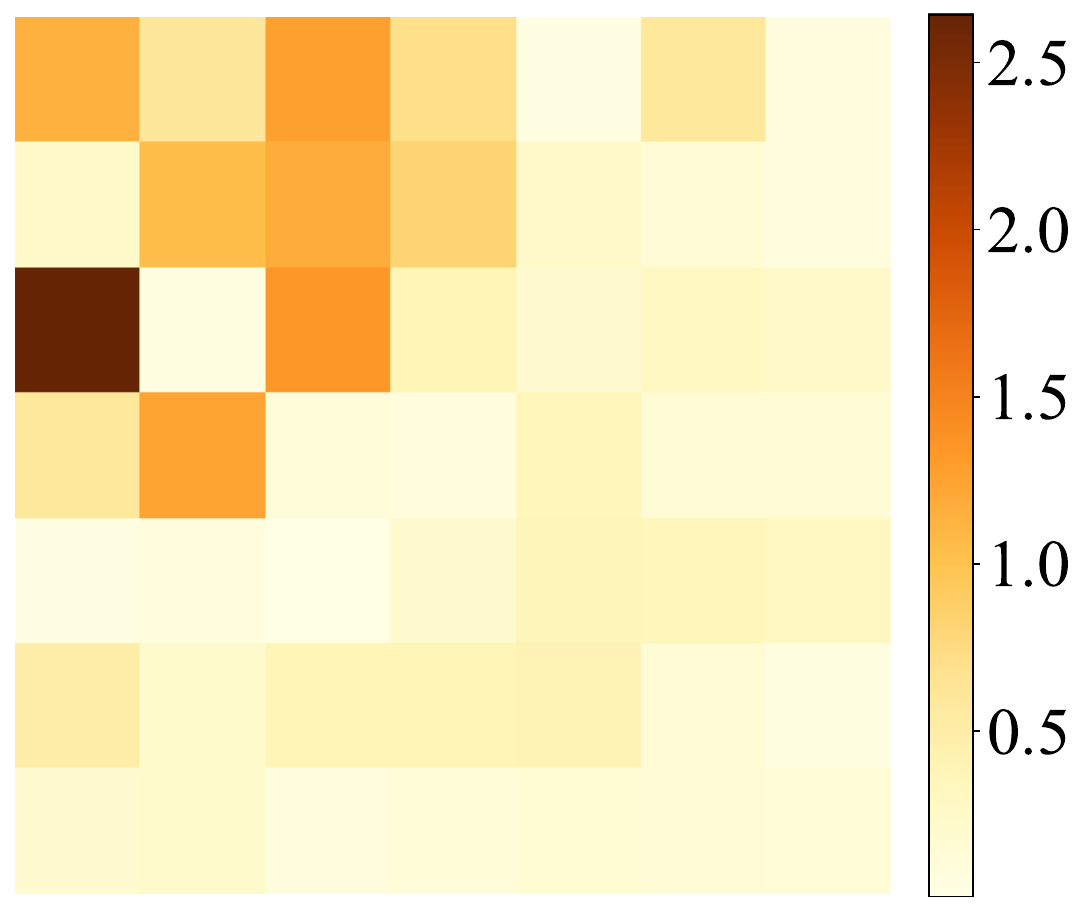}
    \end{subfigure}
    \begin{subfigure}[t]{0.1305\linewidth}
        \centering
        \includegraphics[width=\linewidth]{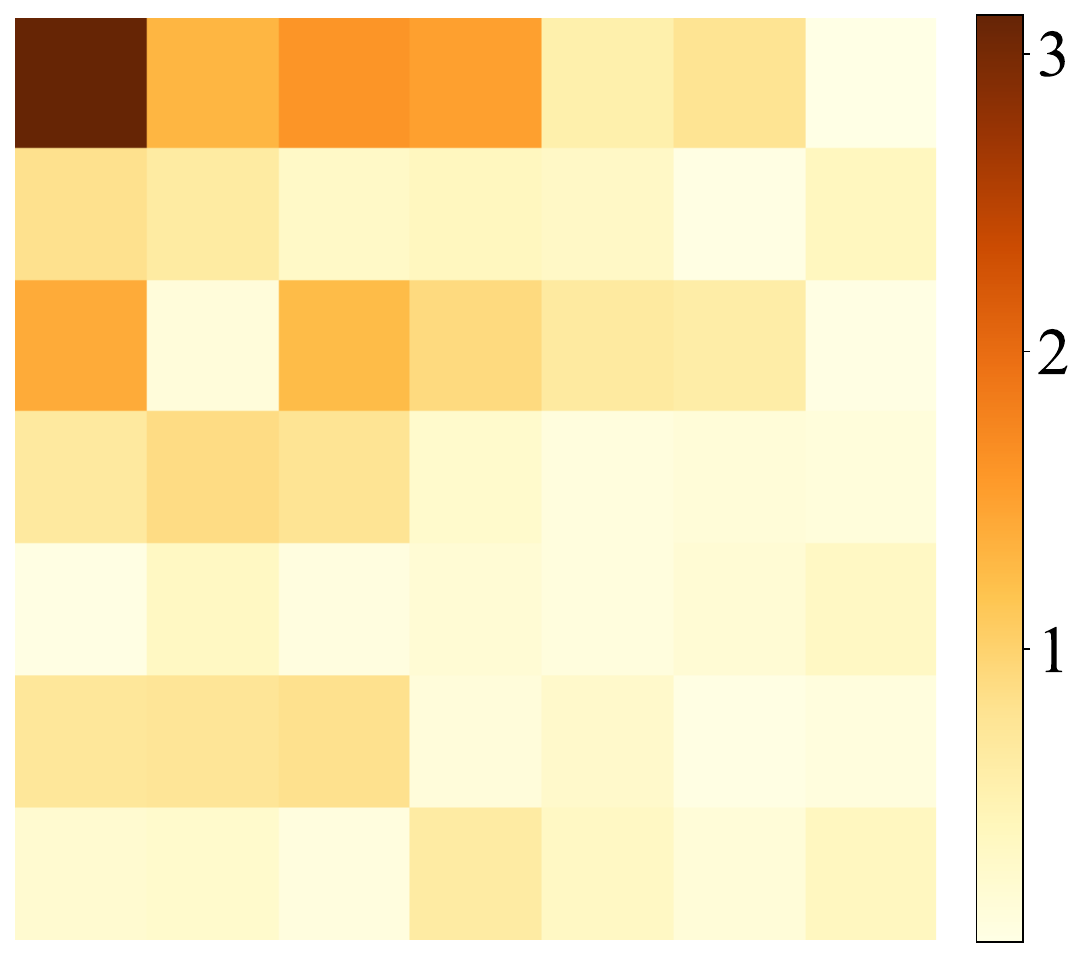}
    \end{subfigure}
    \caption{Visualization of the original feature blocks and their corresponding DCT blocks (absolute values). In the feature domain, the proposed features exhibit higher vertical redundancy, whereas the existing features vary smoothly in both directions. In the DCT domain, the proposed features display more dispersed energy distributions. These distinct redundancy characteristics indicate different coding properties. We use $7\times7$ blocks for $C5$ and $C5$-ReLU and $16\times16$ blocks for other features.}
    \label{fig_org_logdct}
\end{figure*}

\section{Dataset construction and analysis}
\subsection{Dataset construction}
To guarantee the dataset's representativeness and support the long-term study of feature coding research, we curate the test dataset with consideration of three key aspects. 
Table \ref{table_summary} outlines our selected models and tasks, split points, and source data. 
\subsubsection{Model and task selection}
Given the vast number of large models available, it is impractical to include all of them in a single research paper. Therefore, we focus on selecting representative models. To ensure the dataset's representativeness and long-term relevance, we choose one widely recognized model for each type of large model. Specifically, we select DINOv2 \cite{oquab2023dinov2}, Llama3 \cite{dubey2024llama}, and SD3 \cite{esser2024sd3} as representatives of discriminative models, generative models, and hybrid models, respectively.
The trend of adopting Transformer architectures in large models is expected to persist in the coming years. As a result, the selected models are anticipated to retain their research value and practical utility in the near future.
DINOv2 and Llama3 handle visual and textual inputs, respectively, while SD3 processes textual inputs to generate visual outputs. Together, these models provide comprehensive feature representations across visual, textual, and cross-modal applications.

For DINOv2, we include image classification (Cls), semantic segmentation (Seg), and depth estimation (Dpt) tasks. Llama3 is applied to the common sense reasoning task (CSR), and SD3 is used for text-to-image synthesis (TTI). % comment this line to save space.
The selection covers three visual tasks, one textual task, and one text-to-visual task, forming a comprehensive foundation for analyzing feature coding across diverse tasks and model types.

\subsubsection{Split point decision}
The split points are decided to support various downstream tasks. 
For DINOv2, we define two types of split points: 1) \textbf{$SP_{DS}$}, the output of the $40^{th}$ ViT layer, and 2) \textbf{$SP_{DM}$}, which aggregates outputs from the $10^{th}$, $20^{th}$, $30^{th}$, and $40^{th}$ ViT layers. $SP_{DS}$ is used for tasks utilizing single-layer features, while $SP_{DM}$ supports tasks that require multi-layer features. In this study, $SP_{DS}$ is applied to Cls and Seg, and $SP_{DM}$ to Dpt. The corresponding features are designated as $F_{Cls}$, $F_{Seg}$, and $F_{Dpt}$, respectively.

For Llama3, the split point \textbf{$SP_{G}$} is set at the output of the $32^{nd}$ decoder layer. Features extracted at $SP_{G}$ can support both task-specific heads and integration with other large models \cite{wu24next}. In CSR, $SP_{G}$ produces $N \times 4096$ features, where $N$ is the number of tokens. 
% and these features are denoted as $F_{CSR}$.

For SD3, we take the input layer of the VAE decoder as the split point \textbf{$SP_{H}$}. This split point generates features of size $16 \times 128 \times 128$, denoted as $F_{TTI}$, which are then passed to the VAE decoder for image synthesis.
\subsubsection{Source data collection}
% Inspired by the video coding standard VVC \cite{bross2021vvc}, where test sequences are divided into 8 classes with 3–5 video sequences each, 
We organize our feature test dataset into five classes, each supported by a curated subset of publicly available datasets used as source data.
For Cls, we select 500 images from ImageNet \cite{deng2009imagenet}, each representing a unique class accurately classified by DINOv2.
For Seg, we collect 100 images from VOC2012, ensuring coverage of all 20 object classes.
For Dpt, we source 80 images from NYU-Depth-v2, adhering to the train-test split proposed in \cite{lee2019big} and selecting 3-8 images for each scene category.
For CSR, we utilize 500 samples from the Arc-Challenge dataset, chosen for their longest input prompts and correct prediction by Llama3.
For TTI, we first select 500 images from COCO2017 and then collect their longest captions.

% \textcolor{blue}{The selected source data above are used exclusively for the test dataset. Researchers are free to extract training features from their own source data using our open-sourced code. Please find our training dataset in the supplementary.}
% We noticed that learning-based methods may require a dedicated training set. We will provide the source code for generating features from any source data to serve as training data. Please find our training data in the supplementary.
We noticed that learning-based methods require a dedicated training set. Please follow the source code to generate training data and find our training data in the supplementary.
\subsection{Feature analysis}
In this section, we compare the proposed features with those commonly used in existing research to demonstrate the necessity and importance of introducing a new dataset.
\subsubsection{Distribution analysis}
We first compare frequency distributions and cumulative distribution functions (CDFs) of the proposed and commonly used existing features in Fig. \ref{fig_frequency_cdf}. $C5$ and $C5$-ReLU are derived from the official ResNet50 pre-trained on ImageNet, while $P$-layer features are generated by the ResNeXT101-based MaskRCNN pre-trained on COCO2017.

For multi-layer split points, we observe four key distinctions between $SP_{DM}$ and $P$-layer features: (1) $SP_{DM}$ distribution range expands progressively with deeper layers; (2) $SP_{DM}$ distribution is more asymmetric; (3) $SP_{DM}$ feature values are more concentrated, as evidenced by CDFs; and (4) $SP_{DM}$ features contain more regions with sparse feature points. For single-layer split points, $SP_{GS}$ has a more concentrated distribution, while $SP_{H}$ shows more peaks. The proposed dataset omits $C5$-ReLU-like features, as ReLU is rarely used in large models.

The proposed features exhibit different distributions compared to the existing features, highlighting the necessity of introducing the new dataset.
Additionally, the high diversity of the proposed features enhances the dataset's representativeness and suitability for long-term research.

\begin{table}[tbp]
	\footnotesize
	\centering
    \resizebox{0.65\columnwidth}{!}{
	\begin{tabular}{@{}ccc@{}}
		\toprule
		\textbf{Task}                  & \textbf{Head}           & \textbf{Metric}         \\ \midrule
		\textbf{Cls}        & Norm + Linear1          & Accuracy                \\
		\textbf{Seg} & Norm + Linear1          & mIoU                    \\
		\textbf{Dpt}      & Linear4                 & RMSE                    \\ \midrule
		\textbf{CSR}       & Norm + Linear + Softmax & Accuracy                \\ \midrule
		\textbf{TTI}         & VAE Decoder             & FID        \\ \bottomrule
	\end{tabular}}
    \caption{Summary of task head and accuracy metric settings. Find definitions of Heads in the original papers \cite{oquab2023dinov2,dubey2024llama,esser2024sd3}. }
    \label{table_head_metric}
\end{table}

\subsubsection{Redundancy analysis}
Since coding aims to remove redundancy in inputs, we conduct a redundancy comparison on feature blocks. We visualize the original feature blocks and their DCT blocks (absolute values) in Fig. \ref{fig_org_logdct}.
In the feature domain, clear spatial redundancy is observed in both $SP_{DM}$ and $SP_H$ features. $SP_{H}$ exhibits redundancy in both horizontal and vertical directions, while $SP_{DM}$ shows stronger vertical redundancy. 
In contrast, $SP_{GS}$ features vary rapidly in both directions, resulting in weak spatial redundancy. This is because Llama3 only processes textual data, which lacks inherent spatial correlation. 
Existing features, on the other hand, exhibit smoother spatial variation and higher spatial redundancy. 
We attribute this to the translation invariance of convolutional layers, which preserve the relative positions of pixels in the feature domain, resulting in consistent redundancy patterns. In contrast, transformer-based vision models partition images into patches before transforming them, shifting spatial redundancy primarily to the vertical direction, as seen in $SP_{DM}$.

In the DCT domain, spatial redundancy is reflected in the energy distribution. The proposed features display more dispersed energy, whereas existing features concentrate energy in the top-left DC component. $SP_H$ has the highest energy concentration as it closely resembles images.

These distinct redundancy characteristics further demonstrate the necessity of the proposed dataset and suggest that future studies should explore feature coding methods tailored to these unique properties.
\begin{figure}[tp]
	\centering
	\includegraphics[width=\linewidth]{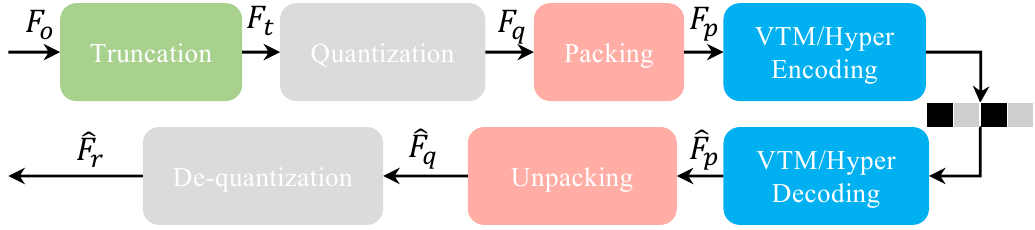}
	\caption{The pipeline of the proposed baseline methods. }
	\label{fig_pipeline}
\end{figure}
\begin{table*}[tbp]
	% \footnotesize
	\centering
    \resizebox{0.95\textwidth}{!}{
	\begin{tabular}{@{}c|ccc|ccc|ccc|ccc|ccc@{}}
    \toprule
    \textbf{Task}                                                                                & \multicolumn{3}{c|}{\textbf{Image Classification}}   & \multicolumn{3}{c|}{\textbf{Semantic Segmentation}} & \multicolumn{3}{c|}{\textbf{Depth Estimation}}     & \multicolumn{3}{c|}{\textbf{Common Sense   Reasoning}} & \multicolumn{3}{c}{\textbf{Text-to-Image   Synthesis}} \\ \midrule
    \textbf{Metric}                                                                              & \textbf{BPFP} & \textbf{Accuracy} & \textbf{MSE}    & \textbf{BPFP}  & \textbf{mIoU}   & \textbf{MSE}    & \textbf{BPFP} & \textbf{RMSE}   & \textbf{MSE}    & \textbf{BPFP}  & \textbf{Accuracy}  & \textbf{MSE}    & \textbf{BPFP}     & \textbf{FID}     & \textbf{MSE}    \\ \midrule
    \textbf{Original}                                                                            & 32             & 100               & 0            & 32              & 83.39           & 0              & 32             & 0.3695        & 0            & 32              & 100                 & 0             & 32                & 0                & 0               \\ \midrule
    % \textbf{Quantization}                                                                        & 10             & 100               & 2.8224       & 10              & 80.87           & 1.7527         & 10             & 0.3885        & 0.5741       & 10              & 100                 & 0.0038        & 10                & 0.07             & 3e-5            \\ \midrule
    \multirow{5}{*}{\textbf{\begin{tabular}[c]{@{}c@{}}VTM \\      Baseline\end{tabular}}}       & 1.94          & 99.80             & 2.9499          & 1.76           & 80.42           & 1.8668          & 2.21          & 0.5021          & 0.6108          & 2.69           & 99.80              & 0.0109          & 1.41              & 6.14             & 0.0066          \\
    & 1.03          & 97.40             & 3.1693          & 0.88           & 79.31           & 2.0836          & 1.27          & 0.6511          & 0.6826          & 1.83           & 100             & 0.0260          & 0.74              & 19.94            & 0.0184          \\
    & 0.23          & 74.60             & 3.6993          & 0.23           & 73.47           & 2.5440          & 0.33          & 0.9530          & 0.9225          & 0.89           & 98.80              & 0.0810          & 0.29              & 58.42            & 0.0421          \\
    & 0.04          & 25.80             & 4.0651          & 0.04           & 56.53           & 2.9526          & 0.03          & 1.4850          & 1.0631          & 0.16           & 82.20              & 0.1868          & 0.11              & 109.23           & 0.0715          \\
    & 0.01          & 7.20              & 4.4400          & 0.01           & 37.42           & 3.2762          & 0.003         & 2.1174          & 1.1072          & 0.04           & 23.80              & 0.2455          & 0.05              & 171.63           & 0.1072          \\ \midrule
    % \textbf{Scaling}                                                                             & 32             & 100               & 3.1677       & 32              & 79.62           & 2.0056         & 32             & 0.3999        & 0.6068       & 32              & 100                 & 0.0038        & 32.00             & 0.03             & 4e-6            \\ \midrule
    \multirow{5}{*}{\textbf{\begin{tabular}[c]{@{}c@{}}Hyperprior\\      Baseline\end{tabular}}} & 2.01          & 93.20             & 3.4501          & 1.71           & 77.96           & 2.2010          & 1.51          & 0.4256          & 0.6600          & 6.34           & 91.40              & 0.0776          & 1.44              & 25.25            & 0.0138          \\
    & 1.13          & 88.80             & 3.6966          & 1.30           & 77.27           & 2.2795          & 1.01          & 0.4965          & 0.6844          & 3.60           & 87.80              & 0.0808          & 0.66              & 52.12            & 0.0300          \\
    & 0.91          & 83.60             & 3.8327          & 0.54           & 74.82           & 2.5845          & 0.43          & 0.6699          & 0.7837          & 1.68           & 82.40              & 0.1622          & 0.28              & 95.03            & 0.0541          \\
    & 0.37          & 29.00             & 4.2909          & 0.12           & 62.58           & 3.0151          & 0.08          & 1.0351          & 1.0844          & 1.50           & 57.80              & 0.1355          & 0.15              & 124.84           & 0.0734          \\
    & 0.23          & 14.60             & 4.7814          & 0.03           & 37.11           & 3.5449          & 0.01          & 2.7302          & 1.1866          & 1.35           & 34.80              & 0.1624          & 0.08              & 170.01           & 0.1006          \\ \bottomrule 
    \end{tabular}}
    \caption{Rate-accuracy (R-A) performance evaluations on the two baseline methods (MSE is calculated between $F_{o}$ and $\hat{F}_{r}$).}
    \label{table_rate_accuracy}
\end{table*}
\section{Unified test conditions}
Even with a test dataset, fair comparisons are still challenging without identical test conditions. To address this, we establish unified test conditions here. 
\subsection{Bitrate computation}
We introduce a new bitrate measurement, Bits Per Feature Point (BPFP), replacing the commonly used Bits Per Pixel (BPP). BPFP is calculated by dividing the total coding bits by the number of feature points. Our rationale for shifting from BPP to BPFP is twofold. First, BPP is unsuitable for features extracted from non-visual data where pixels do not exist. Second, we believe bitrate should reflect the actual encoded data (the feature itself) rather than the source data. Using BPP to calculate bitrate on source data introduces ambiguity and limits fair comparisons between coding methods. For instance, applying the same BPP to different features extracted from the same image can be misleading. In addition, BPP may incentivize downscaling source images to reduce feature size, yielding lower bitrates that do not represent true coding efficiency.
In contrast, BPFP measures bitrate on the feature itself, enabling fair comparisons across coding methods as long as the same feature is used. 

\subsection{Task accuracy evaluation} 
The decoded features undergo task heads to perform accuracy evaluation. 
To isolate the impact of task heads, we choose simple task heads and fix their parameters during evaluation. 
The task heads and accuracy metrics are presented in Table \ref{table_head_metric}. 
% All task head weights are loaded from pre-trained models. 

\begin{table}[tbp]
	\footnotesize
	\centering
    \resizebox{0.98\columnwidth}{!}{
	\begin{tabular}{@{}ccccc@{}}
\toprule
\textbf{Task}                                                                        & \textbf{Split Point} & \textbf{Org. Region}  & \textbf{Trun. R. (VTM)} & \textbf{Trun. R. (Hyperprior)} \\ \midrule
\textbf{Cls}                  & $SP_{DS}$            & {[}-542.30, 94.14{]}  & {[}-20, 20{]}        & {[}-5, 5{]}            \\ \midrule
\textbf{Seg}                  & $SP_{DS}$            & {[}-506.97, 105.95{]} & {[}-20, 20{]}        & {[}-5, 5{]}            \\ \midrule
\multirow{4}{*}{\textbf{Dpt}} & $SP_{DM1}$           & {[}-2.38, 3.27{]}     & {[}-1, 1{]}          & {[}-1, 1{]}            \\
                              & $SP_{DM2}$           & {[}-26.44, 5.03{]}    & {[}-2, 2{]}          & {[}-2, 2{]}            \\
                              & $SP_{DM3}$           & {[}-323.30, 25.05{]}  & {[}-10, 10{]}        & {[}-10, 10{]}          \\
                              & $SP_{DM4}$           & {[}-504.43, 100.27{]} & {[}-20, 20{]}        & {[}-10, 10{]}          \\ \midrule
\textbf{CSR}                  & $SP_{G}$             & {[}-71.50, 47.75{]}   & {[}-5, 5{]}          & {[}-5, 5{]}            \\ \midrule
\textbf{TTI}                  & $SP_{H}$             & {[}-5.79, 4.46{]}     & {[}-5.79, 4.46{]}    & {[}-5.79, 4.46{]}      \\ \bottomrule
\end{tabular}}
    \caption{Summary of the original and truncation region settings.}
    \label{table_trunction}
\end{table}
% \section{\textcolor{blue}{Baseline, benchmark, and insights}}
% 不是为了提供新方法，是为了检验现有方法的可用性。
% insights：还有许多需要做的工作。现有方法的limitation和做不到的。如何advance复杂度。learning方法的通用性。大模型更需要通用性（vtm更实用）
% 特征特别大，深度思考的模型思考也会产生特征，对大数据量的适应性。针对大模型分析编码器的优缺点。
% 单独一节。
\section{Baselines and benchmark}
As a starting point, we introduce two baselines and establish a benchmark. 
% The two baselines are derived from image coding and encompass both handcrafted and learning-based approaches. 
% We adapt these methods to account for the characteristics of large model features. 
% Our goal is to examine their applicability to large model feature coding and provide insights for interested researchers, rather than to develop a completely new method.
We aim to examine their applicability to large model feature coding and provide insights for interested researchers.
\subsection{The baseline methods}
The pipeline for the proposed baselines, illustrated in Fig. \ref{fig_pipeline}, includes pre-processing, core codec, and post-processing modules. First, the original feature $F_{o}$ is truncated to a smaller range and then quantized into integers. Next, the quantized $F_{q}$ is packed into a 2D feature $F_{p}$. The codec takes pre-processed $F_{p}$ as the input and outputs decoded feature $\hat{F}_{p}$. Two core codecs are used: a handcrafted codec VTM and a learning-based codec Hyperprior. In the post-processing module, $\hat{F}_{p}$ is unpacked to $\hat{F}_{q}$ and then converted back into a floating-point feature $\hat{F}_{r}$. 
\subsubsection{Pre-processing and post-processing}
The truncation operation is proposed to remove feature points that deviate significantly from the central region. 
The specific truncated regions for each task are listed in Table \ref{table_trunction}. Different truncations are used for the two baselines for their distinct coding strategies. 
The quantization operation uniformly quantizes the truncated features to 10-bit integers. 
Given the input requirement of VTM, we pack the features into a 2D YUV-400 format. The same packing applies in the Hyperprior baseline. $F_{Seg}$, $F_{Dpt}$, and $F_{TTI}$ are packed into the shapes of $2740 \times 1536$, $3222 \times 6144$, and $512 \times 512$, respectively. Please refer to supplementary materials for the packing details.
The post-processing performs unpacking and de-quantization, which are inverse processes of packing and quantization.
\begin{figure}[tp]
	\centering
	\begin{minipage}{0.49\columnwidth}
		\includegraphics[width=\linewidth]{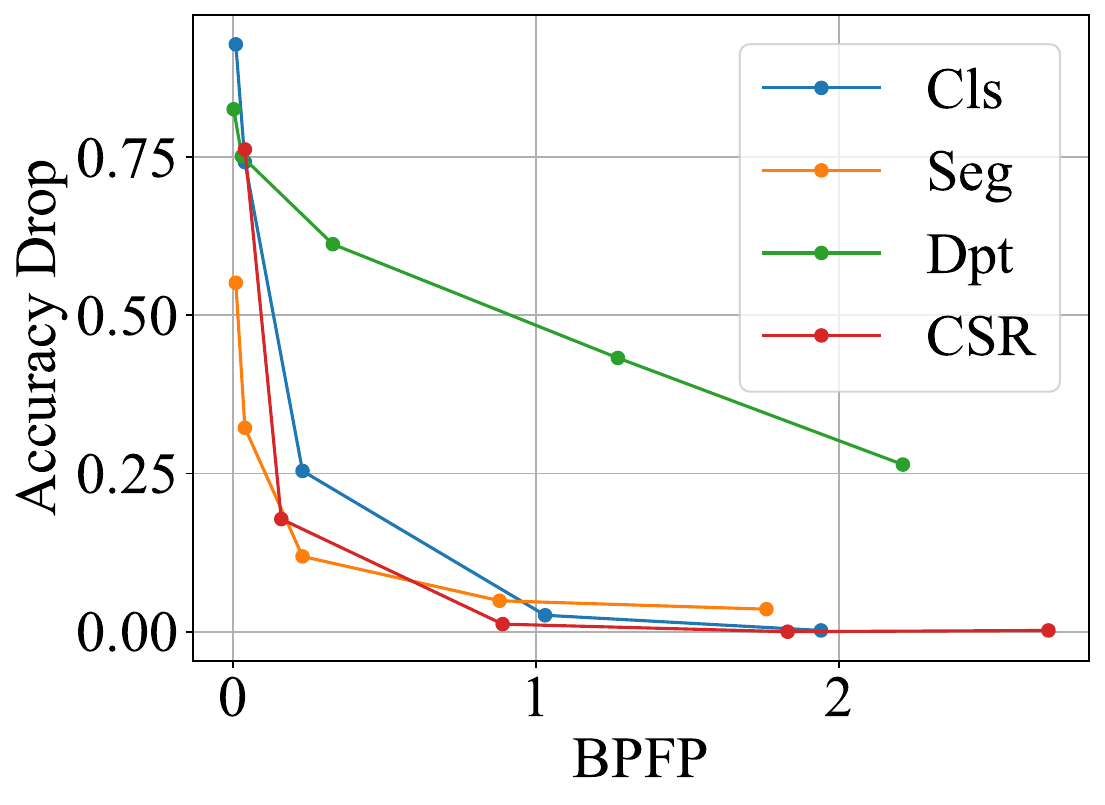}
	\end{minipage}
	\begin{minipage}{0.49\columnwidth}
		\includegraphics[width=\linewidth]{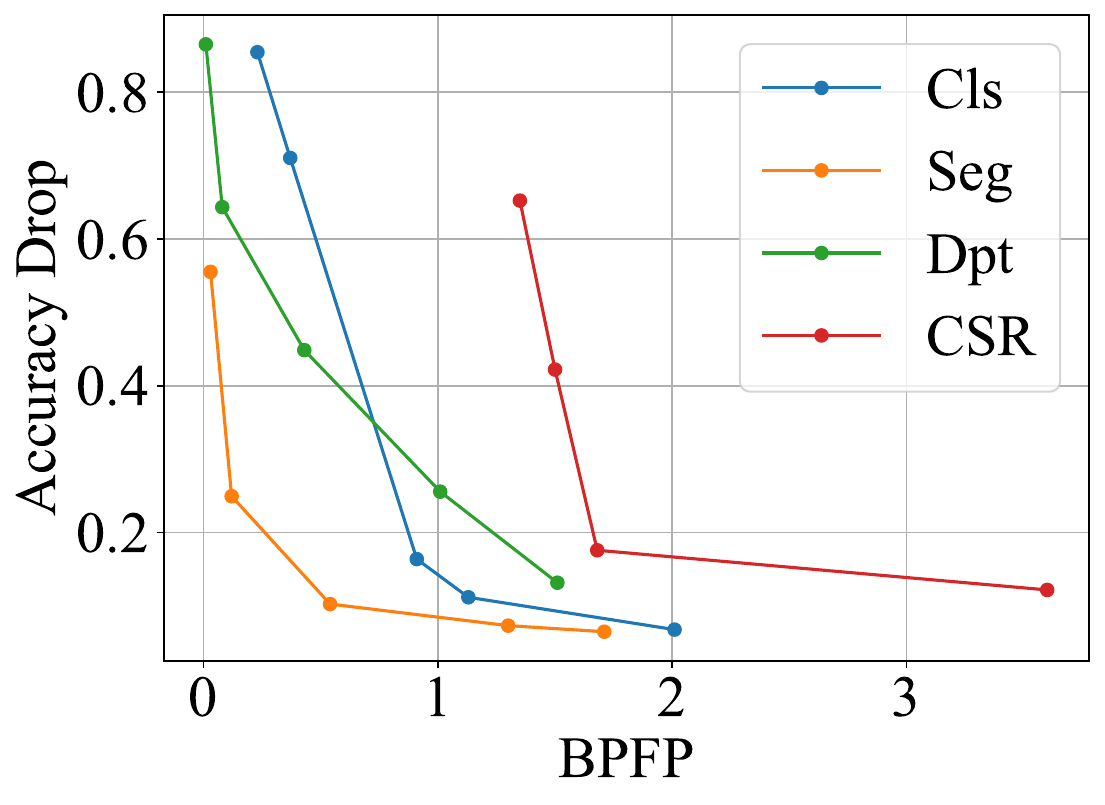}
	\end{minipage}
	\caption{Rate-accuracy-drop (R-AD) comparisons among different tasks. Left: VTM baseline. Right: Hyperprior baseline.}
	\label{fig_rad}
\end{figure}

\subsubsection{Encoding and decoding}
\textbf{VTM baseline:} We employ VTM-23.3 as the codec. The Intra coding is applied using the main configuration file \textit{encoder\_intra\_vtm.cfg}. The \textit{InputChromaFormat} is set to \textit{400}, and \textit{ConformanceWindowMode} is enabled. The \textit{InternalBitDepth}, \textit{InputBitDepth}, and \textit{OutputBitDepth} are all set to \textit{10}. All other configurations are set to default. In our experiments, five QPs \{22, 27, 32, 37, 42\} are used.

% \noindent\textbf{Hyperprior baseline:} We modify the channel of the input and output of the original Hyperprior's input and output layers to accommodate 2D features. The loss function is defined as:
\noindent\textbf{Hyperprior baseline:} We modify the channel of the input and output of the original Hyperprior as 1 to accommodate 2D features. The loss function is defined as:
\begin{equation}
L =  BPFP + \lambda \times ||(F_{o} - \hat{F_{r}})||^2
\end{equation}
where $\lambda$ is a scaling factor used to adjust the bitrate. 
% A larger $\lambda$ corresponds to a higher bitrate.
% In addition, we scale all truncated features into a consistent region of [0,1] before training. 
Please find more details in the supplementary.

\subsection{Rate-accuracy (R-A) analysis}
The R-A evaluation results are presented in Table \ref{table_rate_accuracy}. 
Both baselines can achieve significant bitrate savings, while the VTM baseline outperforms the Hyperprior baseline.
To further assess the impact of bitrate on accuracy, we introduce a new metric, accuracy drop (AD), defined as the percentage decrease in accuracy. For Dpt, the reciprocal of RMSE is employed. The R-AD curves, shown in Fig. \ref{fig_rad}, reveal that most tasks exhibit an inflection point where accuracy starts to decrease significantly. However, for Dpt, this point is observed only in the Hyperprior baseline. In the Hyperprior baseline, inflection points are more distracted across tasks, which may caused by the fact that different models are trained for different tasks. TTI is excluded as images generated from the original features have an FID of 0. 
\subsection{Distortion-accuracy (D-A) analysis}
We evaluate the linear correlation between feature MSE (D) and task accuracy (A) and present coefficients of determination in Table \ref{table_correlation}. Among the baselines, only TTI exhibits high linear correlations. This is reasonable since TTI involves generating images, and the baselines employ image codecs. In contrast, CSR fails to achieve a high linear correlation, as its features are derived from textual data. Overall, the VTM baseline outperforms the Hyperprior baseline, demonstrating its superior adaptability to large model features. Our analysis reveals that feature MSE struggles to characterize semantic distortion, particularly for textual data. This finding suggests that feature MSE is not an effective metric for measuring semantic distortion.
% Developing effective metrics for semantic distortion remains a critical topic in feature coding.
\subsection{Complexity analysis}
We run the VTM baseline and Hyperprior baseline on an Intel® Xeon® E5-2690 v4 CPU and a single NVIDIA GeForce RTX 4090 GPU, respectively. 
The VTM baseline exhibits encoding times ranging from a few seconds to several hours, while its decoding time remains within a few seconds. In contrast, the Hyperprior baseline exhibits comparable encoding and decoding times, most of which are on the order of milliseconds. Detailed results are provided in the supplementary.
We note that runtime depends heavily on both hardware capabilities and implementation details. A well-optimized codec and high-performance hardware can significantly accelerate feature coding. 
% Based on these findings, we recommend using learning-based codecs for real-time applications (e.g., cloud-edge distributed training) and traditional codecs for resource-constrained scenarios (e.g., cloud-centralized training).
\begin{table}[tbp]
	\footnotesize
	\centering
    \resizebox{0.95\columnwidth}{!}{
	\begin{tabular}{@{}cccccc@{}}
    \toprule
    \textbf{Codec}               & \textbf{Cls} & \textbf{Seg} & \textbf{Dpt} & \textbf{CSR} & \textbf{TTI} \\ \midrule
    \textbf{VTM Baseline}        & 0.9413       & 0.8955       & 0.8549       & 0.7490       & 0.9983       \\
    \textbf{Hyperprior Baseline} & 0.9296       & 0.9280       & 0.7437       & 0.4713       & 0.9996       \\ \bottomrule
    \end{tabular}}
    \caption{Distortion-accuracy (D-A) analysis on the two baselines.}
    \label{table_correlation}
\end{table}
\begin{table}[tbp]
	\footnotesize
	\centering
    \resizebox{0.95\columnwidth}{!}{
	\begin{tabular}{@{}c|cc|cc|cc@{}}
    \toprule
    \textbf{Task}                            & \multicolumn{2}{c|}{\textbf{Seg}} & \multicolumn{2}{c|}{\textbf{CSR}} & \multicolumn{2}{c}{\textbf{TTI}} \\ \midrule
    \textbf{Models}                          & \textbf{BPFP}            & \textbf{mIoU}           & \textbf{BPFP}          & \textbf{Acc.}          & \textbf{BPFP}              & \textbf{FID}              \\ \midrule
    \multirow{5}{*}{\textbf{Trained on Seg}} & 1.71                     & 77.96       & 2.08                   & 83.80                      & 0.97                       & 119.83                    \\
    & 1.30                     & 77.27                   & 1.37                   & 66.60                      & 0.62                       & 208.50                    \\
    & 0.54                     & 74.82                   & 0.78                   & 0.20                       & 0.31                       & 333.75                    \\
    & 0.12                     & 62.58                   & 0.25                   & 0.00                       & 0.12                       & 363.32                    \\
    & 0.03                     & 37.11                   & 0.09                   & 0.00                       & 0.05                       & 354.22                    \\ \midrule
    \multirow{5}{*}{\textbf{Trained on TTI}} & 2.58                     & 78.30       & 2.28                   & 97.60                      & 1.44                       & 25.25                     \\
     & 1.96                     & 76.34                   & 1.46                   & 77.80                      & 0.66                       & 52.12                     \\
    & 1.12                     & 73.37                   & 0.65                   & 32.00                      & 0.28                       & 95.03                     \\
    & 0.53                     & 56.90                   & 0.30                   & 0.00                       & 0.15                       & 124.84                    \\
    & 0.26                     & 42.94                   & 0.11                   & 0.00                       & 0.08                       & 170.01                    \\ \midrule
    \multirow{5}{*}{\textbf{Trained on CSR}} & 5.29                     & 78.20       & 6.34                   & 91.40                      & 4.77                       & 58.19                     \\
    & 3.27                     & 77.94                   & 3.60                   & 87.80                      & 2.59                       & 73.19                     \\
    & 1.80                     & 76.77                   & 1.68                   & 82.40                      & 1.02                       & 196.56                    \\
    & 1.73                     & 75.82                   & 1.50                   & 57.80                      & 1.02                       & 191.69                    \\
    & 1.54                     & 75.80                   & 1.35                   & 34.80                      & 0.94                       & 193.08                    \\ \bottomrule
    \end{tabular}}
    \caption{Generalizability evaluation on Seg, CSR, and TTI tasks.}
    \label{table_cross_eval}
    % \vspace{-0.4cm}
\end{table}
\subsection{Generalizability analysis}
We assess the generalizability of trained models by compressing features extracted from one task using models trained on a different task and evaluating their R-A performance.
We select three tasks and present the results in Table \ref{table_cross_eval}. Overall, for a given task (along column), three kinds of models behave differently in both bitrate and accuracy. For example, models trained on CSR struggle to achieve low bitrates when applied to Seg and TTI, while models trained on Seg fail to achieve low FID in TTI. More analyses can be found in the supplementary. In practical applications, an effective compression model should be capable of handling diverse inputs. Therefore, a compression method with high generalizability is highly desirable.
\subsection{Discussion}
Large model applications require a balance of high accuracy, low complexity, and strong generalizability. However, both baselines struggle to maintain high accuracy at low bitrates, underscoring the need for more efficient feature coding strategies tailored to the unique characteristics of large model features. In addition, semantic distortion measurement remains a critical challenge in feature coding. Advancing semantic distortion metrics could further optimize codec performance and enhance overall coding efficiency.

% Each baseline presents distinct trade-offs.
The two baseline approaches present distinct advantages and drawbacks. The handcrafted approach aligns seamlessly with existing codecs but suffers from excessive runtime, limiting its real-time applicability. In contrast, the learning-based approach benefits from efficient GPU acceleration but lacks generalizability, as separate models must be trained for different features. Therefore, developing lightweight and generic codecs remains a promising research direction for feature coding.

\section{Conclusion}
In this paper, we highlight the importance of feature coding in large model deployments and introduce a new research area: large model feature coding. As the first step, we provide a test dataset, unified test conditions, baseline methods, and a benchmark. We examine the applicability of the two baselines and identify promising directions for future research. Our aim is to encourage collaboration between image/video coding and large model communities to drive the development of feature coding.
{
    \small
    \bibliographystyle{ieeenat_fullname}
    \bibliography{refs}
}

\end{document}